\documentclass[aps,prd,preprint,groupedaddress, longbibliography,nofootinbib]{revtex4-2}

\usepackage[utf8]{inputenc}
\usepackage{amsmath,amssymb}
\usepackage{amsfonts}
\usepackage{slashed}
\usepackage{xcolor}

\usepackage[pdftex]{hyperref}
\hypersetup{ 
pdfstartview=FitV,
colorlinks=true, 
linkcolor=black, 
citecolor=blue, 
filecolor=black, 
urlcolor=magenta,
}
\usepackage{graphicx}
\usepackage{changes}

\newcommand{\inc}{\textrm{inc}}

\begin{document}

\preprint{NORDITA 2018-074}

\title{Slow relaxation and diffusion in holographic quantum critical phases}


\author{Richard A. Davison}
\email{davison@damtp.cam.ac.uk}
\affiliation{Department of Physics, Harvard University, Cambridge MA 02138 USA}
\affiliation{Department of Applied Mathematics and Theoretical Physics, University of Cambridge, Cambridge CB3 0WA, United Kingdom}
\author{Simon A. Gentle}
\email{s.a.gentle@uu.nl}
\affiliation{Institute for Theoretical Physics,  Utrecht University, 3508TD Utrecht, The Netherlands}
\affiliation{Instituut-Lorentz for Theoretical Physics, Leiden University,  2333CA Leiden, The Netherlands}
\author{and Blaise Gout\'eraux}
\email{blaise.gouteraux@polytechnique.edu}
\affiliation{Nordita, KTH Royal Institute of Technology and Stockholm University, Roslagstullsbacken 23, SE-106 91 Stockholm, Sweden}

\date{\today}

\begin{abstract}
The dissipative dynamics of strongly interacting systems are often characterised by the timescale set by the inverse temperature $\tau_P\sim\hbar/(k_BT)$. We show that near a class of strongly interacting quantum critical points that arise in the infra-red limit of translationally invariant holographic theories, there is a collective excitation (a quasinormal mode of the dual black hole spacetime) whose lifetime $\tau_{eq}$ is parametrically longer than $\tau_P$: $\tau_{eq}\gg T^{-1}$. The lifetime is enhanced due to its dependence on a dangerously irrelevant coupling that breaks the particle-hole symmetry and the invariance under Lorentz boosts of the quantum critical point. The thermal diffusivity (in units of the butterfly velocity) is anomalously large near the quantum critical point and is governed by $\tau_{eq}$ rather than $\tau_P$. We conjecture that there exists a long-lived, propagating collective mode with velocity $v_s$, and in this case the relation $D=v_s^2\tau_{eq}$ holds exactly in the limit $T\tau_{eq}\gg1$. While scale invariance is broken, a generalised scaling theory still holds provided that the dependence of observables on the dangerously irrelevant coupling is incorporated. Our work further underlines the connection between dangerously irrelevant deformations and slow equilibration.

\end{abstract}


\maketitle

\newpage


In many-body quantum systems with strong interactions, the characteristic timescales relevant for a variety of dynamical processes are short, and are set by the inverse temperature $\tau_P=\hbar/(k_BT)$.\footnote{\baselineskip1pt We will set $\hbar=k_B=1$ from now on.} For example, $\tau_P$ has been shown to control the onset of hydrodynamics in holographic plasmas, the post-quench equilibration of the Sachdev-Ye-Kitaev model, as well as the Lyapunov exponent characterising the growth rate of chaos in both of the aforementioned kinds of theories \cite{Shenker:2013pqa,Kitaevtalk,Maldacena:2015waa,Maldacena:2016hyu,Heller:2016gbp,Eberlein:2017wah}. 
Transport measurements in the strange metallic phase of high-$T_c$ superconductors  \cite{PhysRevLett.59.1337,Zhang:2016ofh} further support the conjecture that $\tau_P$ fundamentally bounds the dynamics of strongly correlated phases \cite{zaanen2004superconductivity,Bruin804,Hartnoll:2014lpa,Gooth,Zaanen:2018edk}.

Indeed, in the vicinity of a quantum critical point (QCP), $T$ is the only energy scale and so the importance of $\tau_P$ is manifest \cite{sachdevbook}. However,  there are circumstances in which non-universal effects are important and lead to dynamics that survive on timescales much longer than $\tau_P$. The most familiar example is near a QCP where translational symmetry is broken by an irrelevant coupling $g$ \cite{PhysRevB.6.1226,PhysRevLett.85.1092,Hartnoll:2012rj,Hartnoll:2014gba,Patel:2014jfa}, leading to the slow relaxation of momentum and a parametrically small resistivity. More generally, whenever the dynamics near a QCP is sensitive to a dangerously irrelevant coupling, $\tau_P$ is no longer privileged since the irrelevant coupling provides an additional energy scale \cite{inbooksachdev}. In such situations, it is not obvious what the relevant timescales for dynamical processes are.

We study a class of strongly interacting, $(d+1)$-dimensional, translationally invariant systems whose  infra-red (IR) physics are governed by hyperscaling violating QCPs with dynamical exponent $z=1$. The particle-hole symmetry and the invariance under Lorentz boosts of the $T=0$ IR QCP are broken by an irrelevant deformation with coupling $g\propto\rho$ the density of the state. We show that in these systems the incoherent current (i.e.~the part of the electric current without momentum drag \cite{Davison:2015taa}) acquires a long lifetime $\tau_{eq}$
\begin{equation}
\label{eq:tauirr}
\tau_{eq}\sim\tau_P\left(\frac{T^{\Delta_g}}{g}\right)^2, 
\end{equation}
which is parametrically longer than $\tau_P$, $\tau_{eq}\gg T^{-1}$, and is controlled by the dimension of the coupling $\Delta_g<0$. While we expect typical excitations to have a lifetime $\tau_P$, it is only after a time $\tau_{eq}$ that local equilibration will be achieved and the expected hydrodynamic behaviour will take over. The slowly relaxing mode produces a narrow peak in the optical conductivity
\begin{equation}
\label{accond}
    \sigma(\omega)=\frac{\rho^2}{sT+\mu\rho}\frac{i}{\omega}+\frac{\sigma_o}{(1-i\omega\tau_{eq})},
\end{equation}
where $\sigma_o$ and $\tau_{eq}$ are given by \eqref{lifetime}, $\rho$ is the charge density, $\mu$ the chemical potential and $s$ the entropy density.
We expect that adding slow momentum relaxation to our theories (as in e.g.~\cite{Gouteraux:2014hca,Davison:2015bea,Blake:2017qgd}) will broaden the divergent $\omega\rightarrow0$ contribution to the conductivity \eqref{accond} into a Drude-like peak. 
The interplay between multiple irrelevant deformations can be subtle but important for transport near QCPs \cite{PhysRevLett.82.4280,PhysRevLett.103.216602,Hartnoll:2011ic,Hartnoll:2014gba,Patel:2014jfa}. 

We study these systems using gauge/gravity duality, where the IR QCP is captured by a spacetime metric that is conformal to AdS$_{d+2}$ and is a solution of Einstein-Dilaton theories with an exponential potential \cite{Charmousis:2010zz,Gouteraux:2011ce}. It is important to note that our models do not capture competing phases on either side of a QCP, only the dynamics of the quantum critical region itself. The irrelevant deformation is realised by a Maxwell field, with exponential coupling to the dilaton, that backreacts on this spacetime and drives a renormalisation group (RG) flow to a non-zero density ultra-violet (UV) fixed point. In gravitational language, we show that certain charged, translationally invariant, asymptotically AdS$_{d+2}$ black branes have quasi-normal modes with parametrically long lifetimes $\sim\tau_{eq}$.

Near the QCP, we furthermore show that $\tau_{eq}$ is the timescale relevant for transport processes that do not involve the dragging of momentum. 
Specifically, at times $t\gtrsim\tau_{eq}$, these processes are diffusive. Near the QCP, they are characterised by a single diffusivity $D_T$ (the thermal diffusivity) where
\begin{equation}
\label{eq:Dirr}
D_T=\frac{2}{d+1-\theta}\;v_B^2 \tau_{eq},
\end{equation}
$\theta$ is a universal number quantifying the violation of hyperscaling at the QCP, and $v_B$ is the `butterfly' velocity at which quantum chaos spreads. The large value of $D_T$ resulting from its sensitivity to irrelevant deformations was established in \cite{Blake:2017qgd}, and was in potential tension with the upper bounds on diffusivities proposed to ensure the causality of diffusive hydrodynamics \cite{Hartman:2017hhp,Lucas:2017ibu}. The result (\ref{eq:Dirr}) elegantly resolves this potential tension: at precisely the timescales at which causality appears to be violated, the diffusive hydrodynamic description breaks down due to the existence of the slowly relaxing mode. This is a consequence of the non-trivial fact that both $D_T$ and $\tau_{eq}$ are governed by the same irrelevant deformation of the QCP.

A number of recent works have established relations similar to (\ref{eq:Dirr}) between thermal diffusivities and the spreading of quantum chaos \cite{Hartnoll:2014lpa,Blake:2016wvh,Blake:2016sud,Gu:2016oyy,Blake:2016jnn,Davison:2016ngz,Patel:2016wdy,Blake:2017qgd,Werman:2017abn,Blake:2017ris}. In holographic theories, these have always been of the form $D_T\sim v_B^2\tau_P$. Our result \eqref{eq:Dirr} lends further support to the claim that in general the timescale appearing in this relation should be $\tau_{eq}$, and not $\tau_P$ or the Lyapunov time $\tau_L$ (which governs the growth rate of quantum chaos) \cite{Hartman:2017hhp,Lucas:2017ibu}. These timescales could not be distinguished in previous examples, which had $\tau_{eq}\sim\tau_L\sim\tau_P$.\footnote{\baselineskip1pt Our holographic systems have the minimum allowed value of the Lyapunov time $\tau_L=\tau_P/(2\pi)$, \cite{Shenker:2013pqa,Maldacena:2015waa}.}
Our results are also non-trivially consistent with the quantum hydrodynamic theory for maximally chaotic systems proposed in \cite{Blake:2017ris} and explored in \cite{Blake:2018leo}. The result $D_T\sim v_B^2\tau_P$ follows from this theory provided that diffusive hydrodynamics applies at timescales $t\sim\tau_P$. Assuming the validity of this theory for the holographic QCPs we study, the parametrically large value of $D_T$ therefore implies that hydrodynamics must break down at timescales $t\sim\tau_{eq}\gg \tau_P$, as we explicitly show.

Another consequence of the additional energy scale $g$ in the IR theory is the violation of naive $\omega/T$ scaling in response functions near the QCP. We close by illustrating this explicitly, and by showing that if one carefully takes into account the $g$-dependence of the critical contribution to the conductivity, a generalised scaling theory \cite{Gouteraux:2013oca,Gouteraux:2014hca,Karch:2014mba, Davison:2018}, which has been applied to dc transport in cuprate strange metals \cite{Hartnoll:2015sea}, continues to hold. Non-trivial scaling theories near QCPs are attractive from a phenomenological point of view: we know that if the strange metallic phase of high $T_c$ superconductors does originate from a QCP, then it cannot be governed by a simple, scale invariant theory, as such a theory is inconsistent with the observed $T$-linear resistivity \cite{2005PhRvL..95j7002P}.

In the remainder of this Letter, we describe our setup and outline the calculations leading to the results mentioned above. We have also found analogous results to \eqref{eq:tauirr} and \eqref{eq:Dirr} in a closely related class of systems that are particle-hole symmetric and flow to QCPs with dangerously irrelevant translational symmetry-breaking deformations. The results for these systems, along with a number of technical details, are presented in \cite{Davison:2018}. In the Appendix, we prove that it is indeed the incoherent current that becomes long-lived at low temperatures, with a lifetime $\tau_{eq}$.

\section{Holographic quantum criticality}

In holographic theories, quantum critical states with dynamical exponent $z=1$ can be described by the $d+2$-dimensional Einstein-Dilaton action \cite{Charmousis:2010zz}
\begin{equation}
\label{actionEMD}
S_{crit}=\int d^{d+2}x \sqrt{-g}\Big(R-\frac12{(\partial\phi)^2}- V_0 e^{-\delta\phi}\Big)\,.
\end{equation}
We study theories where (\ref{actionEMD}) is the effective action capturing the low temperature dynamics far from the boundary of an asymptotically AdS$_{d+2}$ spacetime. By identifying the extra spatial dimension with the energy scale of a dual quantum field theory, (\ref{actionEMD}) describes the IR dynamics that arise at the endpoint of an RG flow generated by deforming a UV CFT. The details of the RG flow will determine the constants $V_0$ and $\delta$, but are otherwise not important for our analysis.

The quantum critical dynamics are captured by the following classical solutions of the action \cite{Charmousis:2010zz,Gouteraux:2011ce}, in which the metric transforms covariantly under the $z=1$ rescaling $(t,\vec x)\mapsto\lambda(t,\vec x)$ 
\begin{equation}
\label{HVsol}
\begin{split}
&d s^2=\left(\frac{u}{L}\right)^{2\frac\theta d-2}\left(-L_t^2d t^2+\tilde L^2 d u^2+L_x^2d\vec{x}^2\right),\quad \tilde L^2=\frac{(d+1-\theta)(d-\theta)}{-V_0},\\
& \phi=\kappa\ln\left(\frac{u}L\right),\quad  \kappa^2=\frac2d\theta(\theta-d)\,,\quad \kappa\delta=2\frac\theta d\,.
\end{split}
\end{equation}
$u$ is the radial coordinate in the IR region of the spacetime $u\gg L$. The running dilaton leads to violation of hyperscaling, parameterised by $\theta<0$ (consistent with the null energy condition). At small temperatures, the entropy density $s\sim T^{d-\theta}$ \cite{Huijse:2011ef} and so the critical state can be thought of as a `CFT' in $(d-\theta)$ spatial dimensions \cite{Kanitscheider:2009as,Gouteraux:2011qh}. $L_t,L_x$ and $L$ are functions of the deformations of the UV fixed point, and depend on the details of the RG flow.  These length scales typically depend smoothly on the scalar source at the boundary (the deformation of the UV CFT) as it is varied over a continuous range of real values. Each such value allows to represent a distinct QCP. From a gravitational perspective these are perhaps better thought of as quantum critical lines \cite{Hartnoll:2011pp,Adam:2012mw,Gouteraux:2012yr,Zaanen:2018edk}.

The RG flow away from the IR critical point produces corrections to the solution (\ref{HVsol}) in inverse powers of $u/L$. For our purposes, the most important correction comes from the Maxwell action
\begin{equation}
\label{actionirrdef}
\Delta S_{irr}=\int d^{d+2}x \sqrt{-g}\frac{Z_0}4 e^{\gamma\phi}F_{\mu\nu}F^{\mu\nu},
\end{equation}
where the constants $Z_0$ and $\gamma$ depend on the details of the flow to the UV fixed point. $\gamma$ encodes the dimension of an irrelevant deformation, as we will shortly illustrate.

Solving the Maxwell equations in the spacetime \eqref{HVsol} gives the profile of the gauge field at leading order in large $u/L$
\begin{equation}
\label{AtsolIR}
    A = A_0 \left(\frac{u}{L}\right)^{\zeta-1} L_t dt\,, \quad  \zeta=d-\kappa\gamma -(d-2)\frac\theta d\,.
\end{equation}
The density $\rho=-Z C^{d/2}A'/\sqrt{BD}\propto A_0$ at $T=0$, so while the gauge field does not backreact on the metric at the QCP, particle-hole symmetry is broken at all temperatures. $A_0$ is the bulk quantity corresponding to the dangerously irrelevant coupling $g$ we referred to in the introduction. Indeed, the gauge field sources corrections to the solution \eqref{HVsol} for the metric and dilaton, which at leading order in $A_0$ are $\sim1+\# A_0^2 u^{2\Delta_{A_0}}$ with $\Delta_{A_0}=(d-\theta+\zeta)/2$. This is an irrelevant deformation if $\Delta_{A_0}<0$ (so the corrections vanish as $u/L\rightarrow\infty$), which we demand from now on. Treating $u$ as an energy scale in the usual way indeed determines the dimension of the irrelevant coupling $A_0$ to be $\Delta_{A_0}$ and that of the corresponding irrelevant operator to be $\Delta_{irr}=d+1-\theta-\Delta_{A_0}$ \cite{Davison:2018}. Therefore, $\Delta_g=\Delta_{A_0}$ in equation \eqref{eq:tauirr}.

\section{Charge response near the QCP}

In order to compute the optical conductivity, we embed the preceding IR theory into a complete holographic RG flow described by the action
\begin{equation}
\label{actionUV}
S=\int d^{d+2}x \sqrt{-g}\Big(R-\frac12{(\partial\phi)^2}-\frac{Z(\phi)}{4}F^2-V(\phi)\Big)\,,
\end{equation}
where $V(\phi)$ and $Z(\phi)$ are chosen to reproduce the IR action \eqref{actionEMD}+\eqref{actionirrdef} as $\phi\rightarrow\infty$. The states we are interested in are captured by the ansatz for the metric $ds^2=-D(r)dt^2+B(r)dr^2+C(r)d\vec{x}^2$, gauge field $A=A(r)dt$ and scalar $\phi=\phi(r)$. $r$ is a radial coordinate that goes to zero at the boundary, where the metric is asymptotically AdS and $A(0)=\mu\ne0$ defines the chemical potential of the state. We are interested in thermal states, and so we assume there is a regular black brane horizon at  $r=r_h>0$, where $D(r\to r_h)=4\pi T(r_h-r)+\ldots$, $B(r\to r_h)=1/(4\pi T(r_h-r))+\ldots$, $C(r\to r_h)=C_h+\ldots$, $\phi(r\to r_h)=\phi_h+\ldots$, $A(r\to r_h)=A_h(r_h-r)+\ldots$. The charge and entropy densities are given by the $r$-independent expressions $\rho=-Z C^{d/2}A'/\sqrt{BD}=Z_h A_h C_h^{d/2}$ and $s=-(\rho A-C^{1+d/2}(D/C)'/\sqrt{BD})/T=4\pi C_h^{d/2}$, where $Z_h\equiv Z(\phi(r_h))$. We are mainly interested in the low $T$ solutions that reduce to \eqref{HVsol} in the IR as $T\rightarrow0$.

The optical conductivity is given by
\begin{equation}
\label{sigmadef}
\sigma(\omega)\equiv-\frac{i}{\omega}\lim_{r\rightarrow0}\left(r^{2-d}\frac{a_x'(r)}{a_x(r)}\right),
\end{equation}
where $a_x$ is the ingoing linear perturbation of the spatial component of the gauge field and obeys the equation \cite{Davison:2015taa}
\begin{equation}
\label{axeq}
\frac{d}{dr}\left[F G\tilde{a}_x'\right]+\omega^2\frac{G}{F}\tilde{a}_x=0\,,
\end{equation}
with $\tilde{a}_x\equiv {a_x}/({sT+\rho A})$, $F\equiv\sqrt{{D}/{B}}$, $G\equiv Z C^{\frac{d}2-1}(sT+\rho A)^2$.

To calculate the low frequency optical conductivity, we use the usual perturbative ansatz \cite{Policastro:2002se}
\begin{equation}
\tilde{a}_x=c\left(\frac{r_h-r}{r_h}\right)^{\frac{-i\omega}{4\pi T}}\left(1+\sum_{n=1}^\infty\left(\frac{i\omega}{4\pi T}\right)^n\mathcal{A}_n(r)\right),
\end{equation}
where $\mathcal{A}_n(r_h)=0$.  Substituting this into \eqref{axeq} and solving at $O(\omega)$ gives
\begin{equation}
\mathcal{A}_1(r)=\int^r_{r_h}d\tilde{r}\left[\alpha\frac{C}{D}\frac{d}{d\tilde{r}}\left(\frac{1}{sT+\rho A}\right)-\frac{1}{r_h-\tilde{r}}\right],
\end{equation}
where $\alpha=s^3T^3Z_h\rho^{-2}(s/4\pi)^{-2/d}$. This results in an optical conductivity \eqref{accond}
where
\begin{equation}
\label{lifetime}
\sigma_o=\frac{s^2T^2 Z_h}{(sT+\rho\mu)^2}\left(\frac{s}{4\pi}\right)^{1-2/d}\,,\quad\tau_{eq}=-\frac{\mathcal{A}_1(0)}{4\pi T}.
\end{equation}
The first term in the optical conductivity \eqref{accond} is the usual small $\omega$ divergence due to momentum conservation, while the second term arises from charge-carrying processes in which no momentum flows \cite{Davison:2015taa}. The pole in the second term at $\omega=-i\tau_{eq}^{-1}$ indicates the existence of a collective excitation with lifetime $\tau_{eq}$. The result \eqref{lifetime} for $\tau_{eq}$ can only be trusted if $\tau_{eq}T\gg 1$, as the perturbative expansion is reliable for $\omega\ll T$.

For a low $T$ state that is sufficiently close to the QCP described by \eqref{HVsol} and \eqref{AtsolIR}, we will now verify that indeed $\tau_{eq}$ is parametrically longer than $T^{-1}$. The deep IR geometry of such a state will have an event horizon at a large value of $u=u_h$, but will still be described by \eqref{HVsol} and \eqref{AtsolIR} over the range $u_{IR}>u>u_{UV}$, with $u_h\gg u_{IR}$ and $u_{UV}\gg L$. Integrating over this part of the spacetime yields a contribution to $\tau_{eq}$ that is independent of the cutoffs \cite{Davison:2018}:
\begin{equation}
\label{lifetimeIR}
\tau_{eq}=\frac{\tilde{L}(d+1-\theta)}{L_tZ_0(1-\zeta)^2}\frac1{A_0^{2}}\left(\frac{u_h}{L}\right)^{1-2\Delta_{A_0}}\sim \frac1{T}\frac{T^{2\Delta_{A_0}}}{A_0^2}\,.
\end{equation}
Recalling that $\Delta_{A_0}<0$, this contribution to $\tau_{eq}$ is parametrically larger than $T^{-1}$ and should dominate the full integral in the limit $T\rightarrow0$. It is manifest that the irrelevant deformation sourced by $A_0$ is responsible for the slow relaxation of the mode, and indeed $\tau_{eq}$ is of the form given in equation \eqref{eq:tauirr} with $g\sim A_0$ and $\Delta_g=\Delta_{A_0}$.

Counterparts of the QCPs \eqref{HVsol} with $z\ne1$ are well-known \cite{Charmousis:2010zz,Gouteraux:2011ce,Huijse:2011ef}. For these solutions, the deformation parameterized by $A_0$ is marginal ($\Delta_{A_0}=0$), and the integral for $\tau_{eq}$ is no longer dominated by the IR spacetime. In these cases we expect $\tau_{eq}\sim1/T$, as has been observed numerically in a variety of holographic theories \cite{Kovtun:2005ev,Edalati:2010hk,Edalati:2010pn,Buchel:2015saa,Janik:2015waa,Buchel:2015ofa,Sybesma:2015oha}.

\section{Diffusivity and hydrodynamics}

As mentioned above, there are two distinct contributions to the small $\omega$ optical conductivity \eqref{accond}. The divergence at $\omega\rightarrow0$ is due to current ($J$) flow that drags (conserved) momentum ($P$), while the remainder is due to current flow that does not. The latter processes can be conveniently isolated by examining the dynamics of the `incoherent' current $J_{\text{inc}}\equiv\chi_{PP}J-\chi_{JP}P$, where $\chi$ denote static susceptibilities \cite{Davison:2015taa}. We will concentrate on $J_{\text{inc}}$: its small $\omega$ conductivity $\sigma_\inc(\omega)$ is proportional to the second term of \eqref{accond}, and is sensitive to the slowly-relaxing mode.\footnote{\baselineskip1pt We restrict to the linear response dynamics around an equilibrium, thermal state.}

Over sufficiently long timescales, we expect relativistic hydrodynamics to govern the system and thus the conductivity of $J_{\text{inc}}$ to be $\omega$-independent \cite{Davison:2015taa}. From \eqref{accond}, it is apparent that this is the case at times $t\gg\tau_{eq}$. In this regime, long wavelength perturbations of $J_{\text{inc}}$ and its associated charge $\delta\rho_{\inc}\equiv s^2T\delta\left(\rho/s\right)$ diffuse with the usual diffusivity $D$ of relativistic hydrodynamics (see e.g.~\cite{Kovtun:2012rj}). $D$ obeys the Einstein relation $D=\sigma_{\text{inc}}^{dc}/\chi_{\text{inc}}$ where $\sigma_{\text{inc}}^{dc}=(sT+\mu\rho)^2\sigma_o$ and $\chi_{\text{inc}}$ is the static susceptibility of $\delta\rho_{\inc}$. While in general $\chi_{\inc}$ depends in a complicated way on the thermodynamic properties of the state, near a QCP it simplifies to $\chi_{\text{inc}}=\rho^2T^2(\partial s/\partial T)_{\rho}$ \cite{Davison:2018}.  Furthermore, as $\sigma_{\text{inc}}^{dc}$ is related to the open-circuit thermal conductivity $\kappa$ by $\sigma_{\text{inc}}^{dc}=T\rho^2\kappa$ in a relativistic hydrodynamic system \cite{Davison:2018}, near the QCP $D$ is equal to the thermal diffusivity $D_T\equiv \kappa/(T\partial s/\partial T)_\rho$. Using our explicit results \eqref{lifetime} for holographic theories, in addition to the temperature scaling of $s$, both diffusivities near the QCP can be written simply as \eqref{eq:Dirr}.

The relation \eqref{eq:Dirr} is possible because $D_T$, $v_B$ and $\tau_{eq}$ are all related to near-horizon properties of the dual black hole.\footnote{\baselineskip1pt The butterfly velocity was computed in terms of the metric near the black hole horizon for the states \eqref{HVsol} in \cite{Roberts:2016wdl,Blake:2016wvh}.} This fact also lies behind the existence of a relation analogous to \eqref{eq:Dirr} for $z\neq1$ QCPs, with $\tau_{eq}$ replaced by $\tau_P$ \cite{Blake:2016sud,Blake:2017qgd}. But unlike in those cases, where $D_T$ and $v_B$ are both properties of the QCP, for the $z=1$ cases at hand the relation \eqref{eq:Dirr} relies crucially on the fact that both $D_T$ and $\tau_{eq}$ depend in the same way on the irrelevant deformation away from the QCP sourced by $A_0$. This is also different to the case of $z=\infty$, $\theta=0$ QCPs, where a relation similar to \eqref{eq:Dirr} with $\tau_{eq}$ replaced by $\tau_P$ arises due to the fact that both $D_T$ and $v_B^2$ are determined by the same irrelevant coupling \cite{Gu:2016oyy,Blake:2016jnn,Davison:2016ngz}.

At times $t\lesssim\tau_{eq}$, relativistic hydrodynamics is not applicable to the system since it doesn't incorporate the dynamics of the slowly relaxing mode that appears at times $t\sim\tau_{eq}$. Since we expect typical excitations near the QCP to have lifetimes $\sim T^{-1}\ll\tau_{eq}$, then it may be possible to identify an effective theory valid to earlier times $t\gtrsim T^{-1}$ by supplementing the hydrodynamic equations to incorporate the existence of the slowly relaxing mode.\footnote{\baselineskip1pt There are other instances in which holographic theories are known to support anomalously long-lived modes with a variety of interesting dispersion relations (typically involving probe branes or higher-derivative actions \cite{Karch:2008fa,Hartnoll:2009ns,Davison:2011ek,Witczak-Krempa:2013aea,Davison:2014lua,Grozdanov:2016vgg,Grozdanov:2016fkt,Chen:2017dsy,Grozdanov:2018fic}).} In the Appendix, we compute holographically the other entries in the matrix of retarded Green's functions for $J$ and $P$ and show they match those of a hydrodynamic theory with a slowly-decaying mode $J_{\text{inc}}$: $\partial_tJ_{\text{inc}}=-J_{\text{inc}}/\tau_{eq}$, using standard techniques \cite{1963AnPhy..24..419K,Hartnoll:2007ih,Kovtun:2012rj}.
Such effective theories typically display pole collisions in the lower half frequency plane, whereby a diffusive mode acquires a real part and turns into a propagating mode at short distances.
The velocity $v_s$ of this propagating mode then determines the diffusivity $D=v_s^2\tau_{eq}$ (see eg (2.17) of \cite{Davison:2014lua}). For \eqref{eq:Dirr} to take this form, we require a velocity $v_s^2=2v_B^2/(d+1-\theta)=1/(d-\theta)$. It is known \cite{Kanitscheider:2009as,Gouteraux:2011qh} that $z=1$, $\theta\neq0$ theories contain a mode with this velocity in their spectrum. We therefore conjecture that this mode of the IR spacetime is promoted to a mode of the full {asymptotically AdS} spacetime,\footnote{\baselineskip1pt This propagating mode will be in addition to the momentum-carrying sound mode, which has a distinct velocity $\sqrt{\partial p/\partial\epsilon}$.} and thus it is because $D=v_s^2\tau_{eq}$ that \eqref{eq:Dirr} is realised. We plan to confirm this picture in  \cite{Davison2019}, using the techniques developed in \cite{Grozdanov:2018fic}.

In light of this discussion, it would be interesting to identify for $z\neq1$ QCPs (where the irrelevant deformation is unimportant) a lifetime $\tau_{eq}\sim1/T$ and velocity $v_s^2\sim v_B^2$ of a collective mode such that $D_T=v_s^2\tau_{eq}$. Such a relation would indicate that it is not the butterfly velocity $v_B$ that fundamentally sets the thermal diffusivity, but instead that \eqref{eq:Dirr} arises due to a relation between the velocities of collective modes and the butterfly velocity near quantum critical points.

\section{Breakdown of $\omega/T$ scaling}

The existence of a collective mode with the parametrically long lifetime $\tau_{eq}$ is the most striking consequence of the breakdown in quantum critical scaling caused by the dangerously irrelevant coupling $A_0$, but it is not the only one. It was previously shown that the conductivity $\sigma_{\inc}(\omega,T)$ does not exhibit $\omega/T$ scaling near the QCPs \eqref{HVsol}: specifically, $\sigma_{\inc}\left(\omega,T=0\right)\sim\omega^{-\zeta}$ \cite{Charmousis:2010zz,Gouteraux:2011ce,Gouteraux:2013oca} while $\sigma_{\inc}^{dc}\sim T^{\zeta+2(d-\theta)}$ \cite{Davison:2015taa}. By carefully keeping track of the dependence on $A_0$ \cite{Davison:2018}, we can explicitly attribute this breakdown in $\omega/T$ scaling to the presence of the irrelevant coupling in the IR theory:

\begin{equation}
\label{sigmaIncscalingz=1}
\sigma_{\inc}^{dc}\sim T^{d-\theta+2\Delta_{A_0}},\;\; \sigma_{\inc}\left(T=0\right)\sim A_0^4\omega^{d-\theta-2\Delta_{A_0}}.
\end{equation}
Recalling that $\Delta_{A_0}=(d-\theta+\zeta)/2$, it is clear that when $z=1$ we can consistently assign $\sigma_{\inc}$ the dimension $\zeta+2(d-\theta)$, and that $\omega/T$ scaling fails because of the non-trivial dependence of $\sigma_{\inc}$ on the irrelevant coupling $A_0$.

In contrast, near the $z\neq1$ counterparts of the QCPs \eqref{HVsol} where $A_0$ sources a marginal deformation $\Delta_{A_{0}}=0$, the incoherent conductivity obeys $\omega/T$ scaling: $\sigma_{\inc}\left(\omega,T=0\right)\sim\omega^{2+(d-2-\theta)/z}$ \cite{Charmousis:2010zz,Gouteraux:2011ce,Gouteraux:2013oca,Gouteraux:2014hca} and $\sigma_{\inc}^{dc}\sim T^{2+(d-2-\theta)/z}$ \cite{Davison:2015taa}.

In both cases ($z=1$ and $z\ne1$), the scaling theory required to account for the total dimension of $\sigma_{\inc}$ is non-trivial. It involves anomalous dimensions for both the entropy density $\Delta_s=d-\theta$ (i.e.~hyperscaling violation) and the charge density $\Delta_\rho=d-\theta+\Phi$ \cite{Gouteraux:2013oca,Gouteraux:2014hca,Karch:2014mba} and is explained in more detail in \cite{Davison:2018}. The anomalous dimension for charge density $\Phi$ is related to the profile of the Maxwell field \eqref{AtsolIR} by $\Phi=(\zeta+\theta-d)/2$, and thus $\Delta_{\rho}=\Delta_{A_0}$ (consistent with our previous observation that $\rho\propto A_0$ at $T=0$). Note that the close relation between $\rho$ and $A_0$, supplemented by a matched asymptotics argument, is at the root of why the charge response near the QCP is sensitive to the irrelevant deformation sourced by $A_0$.

Both the anomalous dimensions, and the extra dimensionful coupling $A_0$, permit a much richer family of $T$-dependence in the quantum critical contribution to the conductivity \eqref{sigmaIncscalingz=1} than is allowed in a simple scale invariant theory \cite{Gouteraux:2014hca,Karch:2014mba}, and may be necessary to explain the various scalings observed in strange metals \cite{PhysRevLett.95.107002,Hartnoll:2015sea}.

\vskip1cm

\begin{acknowledgments}
We would like to thank Sean Hartnoll, Jelle Hartong, Elias Kiritsis and Jan Zaanen for stimulating and insightful discussions. R.A.D. is supported by the Gordon and Betty Moore Foundation Grant GBMF-4306, the STFC Ernest Rutherford Grant ST/R004455/1 and the STFC Consolidated Grant ST/P000681/1. The work of S.A.G. was supported by the Delta-Institute for Theoretical Physics (D-ITP) that is funded by the Dutch Ministry of Education, Culture and Science (OCW). B.G. has been supported during this work by the Marie Curie International Outgoing Fellowship nr 624054 within the 7th European Community Framework Programme FP7/2007-2013, as well as by the European Research Council (ERC) under the European Union's Horizon 2020 research and innovation program (Grant Agreement No341222 and No758759). R.A.D and B.G. wish to thank Nordita for hospitality during the program 'Bounding Transport and Chaos in Condensed Matter and Holography'.
\end{acknowledgments}

\appendix
\section{Conductivity matrix}

In this appendix, we compute the full $2\times2$ matrix of retarded Green's functions for the current $J$ and momentum $P$ operators. From this, we show that the long-lived mode present near $z=1$ IR fixed points is the incoherent current perturbation $J_{\text{inc}}\equiv\chi_{PP}J-\chi_{JP}P$. For simplicity, we work in $d=2$ (four bulk spacetime dimensions). We first compute the matrix holographically, and then compare our result to an effective theory with a conserved incoherent current.

A subtlety arises in the treatment of contact terms, which differ depending on whether the retarded Green's functions are computed using the canonical, Kadanoff-Martin approach \cite{1963AnPhy..24..419K}, or the so-called variational approach (this is the method used in holographic calculations). This is a well-known issue, see for instance section 2 of \cite{Kovtun:2012rj}. It is however meaningful to compare conductivities
\begin{equation}
\label{sigmaAB}
\sigma_{AB}(\omega,q=0)=-\lim_{q\to0}\frac{i}\omega \left[G^R_{AB}(\omega,q)-G^R_{AB}(\omega=0,q)\right]
\end{equation}
which are the response of spatial currents to external sources and do not depend on the choice of contact terms, thanks to the subtraction of the second term in the brackets.

\subsection{Conductivity matrix from holography}

From standard holographic renormalization, the expectation values of the $U(1)$ current and energy-momentum tensor operators are related to the bulk field solutions by
\begin{equation}
\label{Jbdy}
\langle J^\mu\rangle= \lim_{r\to0}\sqrt{-\gamma}Z(\phi)F^{r\mu},
\end{equation}
and 
\begin{equation}
\label{STbdy}
\langle T^{\mu\nu}\rangle=\lim_{r\to0}\frac2{r^5}\left[K^{\mu\nu}-K\gamma^{\mu\nu}+G_{(\gamma)}^{\mu\nu}-\left(2+\frac{(3-\Delta)}4\phi^2+\frac14(\nabla_{(\gamma)}\phi)^2\right)\gamma^{\mu\nu}-\frac12\nabla^\mu_{(\gamma)}\nabla^\nu_{(\gamma)}\phi\right],
\end{equation}
respectively. We work in radial gauge ($g_{rx}=0$) and study time-dependent perturbations $a_x=a_x(r)e^{-i\omega t}$, $g_{t}^x= g_{t}^x(r)e^{-i\omega t}$ with the following expansions near the AdS boundary of the spacetime
\begin{equation}
\label{UVexpFluc}
\begin{split}
 a_x & =  a^{(0)}+ a^{(1)} r+O(r^2),\\
 g_{t}^x &=  g^{(0)}+ g^{(3)}r^3+O(r^4).
\end{split}
\end{equation}
Plugging \eqref{UVexpFluc} into \eqref{Jbdy} and \eqref{STbdy}, we find that the expectation values of $J$ and $P$ are related to the bulk field solutions by
\begin{equation}
\label{JbdyOS}
\langle  J\rangle=  a^{(1)}-\rho  g^{(0)}\,,\qquad \langle  P\rangle=-3 g^{(3)}-\epsilon g^{(0)}\,,
\end{equation}
where $\epsilon$ is the energy density.
The equations of motion impose the following condition on the solution
\begin{equation}
3 g^{(3)}=\rho a^{(0)},
\end{equation}
where $\rho$ is the background charge density. Thus, we find that the expectation values of $J$ and $P$ are related to the bulk field solutions by
\begin{equation}
\label{JbdyOS}
\langle  J\rangle=  a^{(1)}-\rho  g^{(0)}\,,\qquad \langle  P\rangle=-\rho a^{(0)}-\epsilon g^{(0)}\,.
\end{equation}
In the main text, we have solved for the entire bulk fluctuation $a_x(r)$ at small frequencies. Expanding the solution close to the boundary, we can express $a^{(1)}$ as a function of $a^{(0)}$, and therefore the expectation values of $J$ and $P$ in terms of the sources $a^{(0)}$ and $g^{(0)}$. From these expressions, we can directly read off the $(J,P)$ retarded Green's function matrix:
\begin{equation}
\begin{split}
&G^R_{PP}(\omega,q=0)=-\epsilon\,,\\
& G^R_{PJ}(\omega,q=0)= G^R_{JP}(\omega,q=0)=-\rho\,,\\
& G^R_{JJ}(\omega,q=0)=-\frac{\rho^2}{sT+\mu\rho}+\frac{i\omega\sigma_o}{(1-i\omega\tau_{eq})}.
\end{split}
\end{equation}

To compute the conductivity matrix we are interested in, we must compute the second term in brackets of \eqref{sigmaAB}. To do this, we solve for bulk fluctuations $\delta a_x$, $\delta g_{t}^x$ at zero frequencies and wavevectors. The system of bulk perturbation equations is solved by
\begin{equation}
\label{staticsol}
\delta a_x(r)=-A(r)\,,\qquad \delta g_{t}^x=\frac{D(r)}{C(r)}
\end{equation}
from which we read off
\begin{equation}
a^{(0)}=-\mu\,,\qquad a^{(1)}=\rho\,,\qquad g^{(0)}=1\,,\qquad g^{(3)}=-\frac13(\epsilon+p)
\end{equation}
Plugging \eqref{staticsol} into \eqref{Jbdy}, \eqref{STbdy} returns
\begin{equation}
\langle  J\rangle_{\omega=0}=0\,,\qquad \langle  P\rangle_{\omega=0}=p\,.
\end{equation}
This leads to 
\begin{equation}
G_{PP}(\omega=0,q\to0)=p
\end{equation}
and all other elements of the matrix are zero.

Putting everything together into \eqref{sigmaAB},we find the expression (2) in the main text for the charge conductivity which we recall here for convenience
\begin{equation}
\label{accondSM}
    \sigma_{JJ}(\omega)=\frac{\rho^2}{sT+\mu\rho}\frac{i}{\omega}+\frac{\sigma_o}{(1-i\omega\tau_{eq})},
\end{equation}
in addition to the other elements of the conductivity matrix
\begin{equation}
\label{OtherSigma}
\sigma_{JP}=\sigma_{PJ}=\rho\frac{i}{\omega}\,,\quad \sigma_{PP}=\chi_{PP}\frac{i}{\omega},
\end{equation}
where the static susceptibilities
\begin{equation}
\chi_{JP}=\chi_{PJ}=\rho\,,\qquad \chi_{PP}=\epsilon+p=sT+\mu\rho\,.
\end{equation}

Upon changing basis from $(J,P)$ to $(J_{\text{inc}},P)$, we find that the conductivity matrix diagonalises with non-vanishing elements
\begin{equation}
\label{eq:diagonalisedconds}
\sigma_{J_{\inc}J_{\inc}}= \frac{\chi_{PP}^2\sigma_o}{(1-i\omega\tau_{eq})},\quad\quad\quad\sigma_{PP}=\chi_{PP}\frac{i}{\omega}.
\end{equation}
The diagonalisation means that the dynamics of small perturbations $J_{\inc}$ and $P$ are completely independent at late times. Furthermore, they are manifestly controlled by modes with two different lifetimes. Momentum $P$ relaxes at a rate $\tau^{-1}=0$ (i.e.~it is conserved), while the incoherent current $J_{\inc}$ relaxes at a rate $\tau^{-1}=\tau_{eq}^{-1}$. This unambiguously shows that the slowly relaxing mode of the system is indeed $J_{\inc}$.

\subsection{Conductivity matrix from an effective theory with an almost conserved incoherent current}

To emphasize this point, we can uniquely identify the effective theory governing the late-time dynamics of a quantum field theory with the properties we have just described. The effective theory must have the form
\begin{equation}
\label{eq:hydroansatz}
\partial_t\langle P\rangle=0,\quad\quad\quad \partial_t \langle\tilde{J}\rangle=-\tau_{eq}^{-1}\tilde{J},
\end{equation}
where the first equation is simply the conservation of total momentum. The second equation imposes slow relaxation of an operator $\tilde{J}\equiv J+cP$ that overlaps with the current $J$, to ensure that the electrical conductivity has a pole with a long lifetime $\tau_{eq}$. Note that we can write this second equation as diagonal in $\tilde{J}$ without loss of generality, due to the conservation equation for momentum.

Computing the conductivities that follow from this class of effective theories (using the standard techniques \cite{1963AnPhy..24..419K,Hartnoll:2007ih}), we find that
\begin{equation}
\begin{aligned}
&\sigma_{PP}=\chi_{PP}\frac{i}{\omega},\quad\quad \sigma_{JP}=\frac{i}{\omega}\left[\chi_{JP}-\frac{\chi_{JP}+c\chi_{PP}}{1-i\omega\tau_{eq}}\right],\\
&\sigma_{PJ}=\chi_{PJ}\frac{i}{\omega},\quad\quad \sigma_{JJ}=-c\chi_{JP}\frac{i}{\omega}+\frac{\tau_{eq}\left(c\chi_{JP}+\chi_{JJ}\right)}{1-i\omega\tau_{eq}}.
\end{aligned}
\end{equation}
Consistency between these effective theory results and the holographic expressions \eqref{accondSM} and \eqref{OtherSigma} requires setting $c=-\chi_{JP}/\chi_{PP}$. This condition is equivalent to $\tilde{J}=J_\inc$ i.e.~it is the expectation value of the incoherent current that obeys the almost-conservation equation \eqref{eq:hydroansatz}.

We find
\begin{equation}
\sigma_{JJ}= \frac{\det\chi\tau_{eq}}{\chi_{PP}}\frac{1}{(1-i\omega\tau_{eq})}+\frac{\chi_{JP}^2}{\chi_{PP}}\frac{i}{\omega}.
\end{equation}
from which we identify
\begin{equation}
\sigma_o=\frac{\det\chi\tau_{eq}}{\chi_{PP}}\,.
\end{equation}

\bibliography{incoherent-biblio}

\begin{thebibliography}{74}%
\makeatletter
\providecommand \@ifxundefined [1]{%
 \@ifx{#1\undefined}
}%
\providecommand \@ifnum [1]{%
 \ifnum #1\expandafter \@firstoftwo
 \else \expandafter \@secondoftwo
 \fi
}%
\providecommand \@ifx [1]{%
 \ifx #1\expandafter \@firstoftwo
 \else \expandafter \@secondoftwo
 \fi
}%
\providecommand \natexlab [1]{#1}%
\providecommand \enquote  [1]{``#1''}%
\providecommand \bibnamefont  [1]{#1}%
\providecommand \bibfnamefont [1]{#1}%
\providecommand \citenamefont [1]{#1}%
\providecommand \href@noop [0]{\@secondoftwo}%
\providecommand \href [0]{\begingroup \@sanitize@url \@href}%
\providecommand \@href[1]{\@@startlink{#1}\@@href}%
\providecommand \@@href[1]{\endgroup#1\@@endlink}%
\providecommand \@sanitize@url [0]{\catcode `\\12\catcode `\$12\catcode
  `\&12\catcode `\#12\catcode `\^12\catcode `\_12\catcode `\%12\relax}%
\providecommand \@@startlink[1]{}%
\providecommand \@@endlink[0]{}%
\providecommand \url  [0]{\begingroup\@sanitize@url \@url }%
\providecommand \@url [1]{\endgroup\@href {#1}{\urlprefix }}%
\providecommand \urlprefix  [0]{URL }%
\providecommand \Eprint [0]{\href }%
\providecommand \doibase [0]{http://dx.doi.org/}%
\providecommand \selectlanguage [0]{\@gobble}%
\providecommand \bibinfo  [0]{\@secondoftwo}%
\providecommand \bibfield  [0]{\@secondoftwo}%
\providecommand \translation [1]{[#1]}%
\providecommand \BibitemOpen [0]{}%
\providecommand \bibitemStop [0]{}%
\providecommand \bibitemNoStop [0]{.\EOS\space}%
\providecommand \EOS [0]{\spacefactor3000\relax}%
\providecommand \BibitemShut  [1]{\csname bibitem#1\endcsname}%
\let\auto@bib@innerbib\@empty
\bibitem [{\citenamefont {Shenker}\ and\ \citenamefont
  {Stanford}(2014)}]{Shenker:2013pqa}%
  \BibitemOpen
  \bibfield  {author} {\bibinfo {author} {\bibfnamefont {Stephen~H.}\
  \bibnamefont {Shenker}}\ and\ \bibinfo {author} {\bibfnamefont {Douglas}\
  \bibnamefont {Stanford}},\ }\bibfield  {title} {\enquote {\bibinfo {title}
  {{Black holes and the butterfly effect}},}\ }\href {\doibase
  10.1007/JHEP03(2014)067} {\bibfield  {journal} {\bibinfo  {journal} {JHEP}\
  }\textbf {\bibinfo {volume} {03}},\ \bibinfo {pages} {067} (\bibinfo {year}
  {2014})},\ \Eprint {http://arxiv.org/abs/1306.0622} {arXiv:1306.0622
  [hep-th]} \BibitemShut {NoStop}%
\bibitem [{\citenamefont {Kitaev}(2015)}]{Kitaevtalk}%
  \BibitemOpen
  \bibfield  {author} {\bibinfo {author} {\bibfnamefont {Alexei}\ \bibnamefont
  {Kitaev}},\ }\bibfield  {title} {\enquote {\bibinfo {title} {{Talks at
  KITP}},}\ }\href@noop {} {\bibfield  {journal} {\bibinfo  {journal} {Talks at
  `Entanglement in Strongly-Correlated Quantum Matter', KITP}\ } (\bibinfo
  {year} {2015})}\BibitemShut {NoStop}%
\bibitem [{\citenamefont {Maldacena}\ \emph {et~al.}(2016)\citenamefont
  {Maldacena}, \citenamefont {Shenker},\ and\ \citenamefont
  {Stanford}}]{Maldacena:2015waa}%
  \BibitemOpen
  \bibfield  {author} {\bibinfo {author} {\bibfnamefont {Juan}\ \bibnamefont
  {Maldacena}}, \bibinfo {author} {\bibfnamefont {Stephen~H.}\ \bibnamefont
  {Shenker}}, \ and\ \bibinfo {author} {\bibfnamefont {Douglas}\ \bibnamefont
  {Stanford}},\ }\bibfield  {title} {\enquote {\bibinfo {title} {{A bound on
  chaos}},}\ }\href {\doibase 10.1007/JHEP08(2016)106} {\bibfield  {journal}
  {\bibinfo  {journal} {JHEP}\ }\textbf {\bibinfo {volume} {08}},\ \bibinfo
  {pages} {106} (\bibinfo {year} {2016})},\ \Eprint
  {http://arxiv.org/abs/1503.01409} {arXiv:1503.01409 [hep-th]} \BibitemShut
  {NoStop}%
\bibitem [{\citenamefont {Maldacena}\ and\ \citenamefont
  {Stanford}(2016)}]{Maldacena:2016hyu}%
  \BibitemOpen
  \bibfield  {author} {\bibinfo {author} {\bibfnamefont {Juan}\ \bibnamefont
  {Maldacena}}\ and\ \bibinfo {author} {\bibfnamefont {Douglas}\ \bibnamefont
  {Stanford}},\ }\bibfield  {title} {\enquote {\bibinfo {title} {{Remarks on
  the Sachdev-Ye-Kitaev model}},}\ }\href {\doibase 10.1103/PhysRevD.94.106002}
  {\bibfield  {journal} {\bibinfo  {journal} {Phys. Rev.}\ }\textbf {\bibinfo
  {volume} {D94}},\ \bibinfo {pages} {106002} (\bibinfo {year} {2016})},\
  \Eprint {http://arxiv.org/abs/1604.07818} {arXiv:1604.07818 [hep-th]}
  \BibitemShut {NoStop}%
\bibitem [{\citenamefont {Heller}(2016)}]{Heller:2016gbp}%
  \BibitemOpen
  \bibfield  {author} {\bibinfo {author} {\bibfnamefont {Michal~P.}\
  \bibnamefont {Heller}},\ }\bibfield  {title} {\enquote {\bibinfo {title}
  {{Holography, Hydrodynamization and Heavy-Ion Collisions}},}\ }\bibfield
  {booktitle} {\emph {\bibinfo {booktitle} {{Proceedings, 56th Cracow School of
  Theoretical Physics : A Panorama of Holography: Zakopane, Poland, May 24-June
  1, 2016}}},\ }\href {\doibase 10.5506/APhysPolB.47.2581} {\bibfield
  {journal} {\bibinfo  {journal} {Acta Phys. Polon.}\ }\textbf {\bibinfo
  {volume} {B47}},\ \bibinfo {pages} {2581} (\bibinfo {year} {2016})},\ \Eprint
  {http://arxiv.org/abs/1610.02023} {arXiv:1610.02023 [hep-th]} \BibitemShut
  {NoStop}%
\bibitem [{\citenamefont {Eberlein}\ \emph {et~al.}(2017)\citenamefont
  {Eberlein}, \citenamefont {Kasper}, \citenamefont {Sachdev},\ and\
  \citenamefont {Steinberg}}]{Eberlein:2017wah}%
  \BibitemOpen
  \bibfield  {author} {\bibinfo {author} {\bibfnamefont {Andreas}\ \bibnamefont
  {Eberlein}}, \bibinfo {author} {\bibfnamefont {Valentin}\ \bibnamefont
  {Kasper}}, \bibinfo {author} {\bibfnamefont {Subir}\ \bibnamefont {Sachdev}},
  \ and\ \bibinfo {author} {\bibfnamefont {Julia}\ \bibnamefont {Steinberg}},\
  }\bibfield  {title} {\enquote {\bibinfo {title} {{Quantum quench of the
  Sachdev-Ye-Kitaev Model}},}\ }\href {\doibase 10.1103/PhysRevB.96.205123}
  {\bibfield  {journal} {\bibinfo  {journal} {Phys. Rev.}\ }\textbf {\bibinfo
  {volume} {B96}},\ \bibinfo {pages} {205123} (\bibinfo {year} {2017})},\
  \Eprint {http://arxiv.org/abs/1706.07803} {arXiv:1706.07803
  [cond-mat.str-el]} \BibitemShut {NoStop}%
\bibitem [{\citenamefont {Gurvitch}\ and\ \citenamefont
  {Fiory}(1987)}]{PhysRevLett.59.1337}%
  \BibitemOpen
  \bibfield  {author} {\bibinfo {author} {\bibfnamefont {M.}~\bibnamefont
  {Gurvitch}}\ and\ \bibinfo {author} {\bibfnamefont {A.~T.}\ \bibnamefont
  {Fiory}},\ }\bibfield  {title} {\enquote {\bibinfo {title} {Resistivity of
  ${\mathrm{la}}_{1.825}$${\mathrm{sr}}_{0.175}$${\mathrm{cuo}}_{4}$ and
  ${\mathrm{yba}}_{2}$${\mathrm{cu}}_{3}$${\mathrm{o}}_{7}$ to 1100 k: Absence
  of saturation and its implications},}\ }\href {\doibase
  10.1103/PhysRevLett.59.1337} {\bibfield  {journal} {\bibinfo  {journal}
  {Phys. Rev. Lett.}\ }\textbf {\bibinfo {volume} {59}},\ \bibinfo {pages}
  {1337--1340} (\bibinfo {year} {1987})}\BibitemShut {NoStop}%
\bibitem [{\citenamefont {Zhang}\ \emph {et~al.}(2017)\citenamefont {Zhang},
  \citenamefont {Levenson-Falk}, \citenamefont {Ramshaw}, \citenamefont {Bonn},
  \citenamefont {Liang}, \citenamefont {Hardy}, \citenamefont {Hartnoll},\ and\
  \citenamefont {Kapitulnik}}]{Zhang:2016ofh}%
  \BibitemOpen
  \bibfield  {author} {\bibinfo {author} {\bibfnamefont {J.~C.}\ \bibnamefont
  {Zhang}}, \bibinfo {author} {\bibfnamefont {E.~M.}\ \bibnamefont
  {Levenson-Falk}}, \bibinfo {author} {\bibfnamefont {B.~J.}\ \bibnamefont
  {Ramshaw}}, \bibinfo {author} {\bibfnamefont {D.~A.}\ \bibnamefont {Bonn}},
  \bibinfo {author} {\bibfnamefont {R.}~\bibnamefont {Liang}}, \bibinfo
  {author} {\bibfnamefont {W.~N.}\ \bibnamefont {Hardy}}, \bibinfo {author}
  {\bibfnamefont {S.~A.}\ \bibnamefont {Hartnoll}}, \ and\ \bibinfo {author}
  {\bibfnamefont {A.}~\bibnamefont {Kapitulnik}},\ }\bibfield  {title}
  {\enquote {\bibinfo {title} {{Anomalous Thermal Diffusivity in Underdoped
  YBa$_2$Cu$_3$O$_{6+x}$}},}\ }\href {\doibase 10.1073/pnas.1703416114}
  {\bibfield  {journal} {\bibinfo  {journal} {Proc. Nat. Acad. Sci.}\ }\textbf
  {\bibinfo {volume} {114}},\ \bibinfo {pages} {5378--5383} (\bibinfo {year}
  {2017})},\ \Eprint {http://arxiv.org/abs/1610.05845} {arXiv:1610.05845
  [cond-mat.supr-con]} \BibitemShut {NoStop}%
\bibitem [{\citenamefont {Zaanen}(2004)}]{zaanen2004superconductivity}%
  \BibitemOpen
  \bibfield  {author} {\bibinfo {author} {\bibfnamefont {Jan}\ \bibnamefont
  {Zaanen}},\ }\bibfield  {title} {\enquote {\bibinfo {title}
  {Superconductivity: Why the temperature is high},}\ }\href@noop {} {\bibfield
   {journal} {\bibinfo  {journal} {Nature}\ }\textbf {\bibinfo {volume}
  {430}},\ \bibinfo {pages} {512--513} (\bibinfo {year} {2004})}\BibitemShut
  {NoStop}%
\bibitem [{\citenamefont {Bruin}\ \emph {et~al.}(2013)\citenamefont {Bruin},
  \citenamefont {Sakai}, \citenamefont {Perry},\ and\ \citenamefont
  {Mackenzie}}]{Bruin804}%
  \BibitemOpen
  \bibfield  {author} {\bibinfo {author} {\bibfnamefont {J.~A.~N.}\
  \bibnamefont {Bruin}}, \bibinfo {author} {\bibfnamefont {H.}~\bibnamefont
  {Sakai}}, \bibinfo {author} {\bibfnamefont {R.~S.}\ \bibnamefont {Perry}}, \
  and\ \bibinfo {author} {\bibfnamefont {A.~P.}\ \bibnamefont {Mackenzie}},\
  }\bibfield  {title} {\enquote {\bibinfo {title} {Similarity of scattering
  rates in metals showing t-linear resistivity},}\ }\href {\doibase
  10.1126/science.1227612} {\bibfield  {journal} {\bibinfo  {journal}
  {Science}\ }\textbf {\bibinfo {volume} {339}},\ \bibinfo {pages} {804--807}
  (\bibinfo {year} {2013})}\BibitemShut {NoStop}%
\bibitem [{\citenamefont {Hartnoll}(2015)}]{Hartnoll:2014lpa}%
  \BibitemOpen
  \bibfield  {author} {\bibinfo {author} {\bibfnamefont {Sean~A.}\ \bibnamefont
  {Hartnoll}},\ }\bibfield  {title} {\enquote {\bibinfo {title} {{Theory of
  universal incoherent metallic transport}},}\ }\href {\doibase
  10.1038/nphys3174} {\bibfield  {journal} {\bibinfo  {journal} {Nature Phys.}\
  }\textbf {\bibinfo {volume} {11}},\ \bibinfo {pages} {54} (\bibinfo {year}
  {2015})},\ \Eprint {http://arxiv.org/abs/1405.3651} {arXiv:1405.3651
  [cond-mat.str-el]} \BibitemShut {NoStop}%
\bibitem [{\citenamefont {Gooth}\ \emph {et~al.}(2018)\citenamefont {Gooth},
  \citenamefont {Menges}, \citenamefont {Kumar}, \citenamefont {S{\"u}b},
  \citenamefont {Shekhar}, \citenamefont {Sun}, \citenamefont {Drechsler},
  \citenamefont {Zierold}, \citenamefont {Felser},\ and\ \citenamefont
  {Gotsmann}}]{Gooth}%
  \BibitemOpen
  \bibfield  {author} {\bibinfo {author} {\bibfnamefont {J.}~\bibnamefont
  {Gooth}}, \bibinfo {author} {\bibfnamefont {F.}~\bibnamefont {Menges}},
  \bibinfo {author} {\bibfnamefont {N.}~\bibnamefont {Kumar}}, \bibinfo
  {author} {\bibfnamefont {V.}~\bibnamefont {S{\"u}b}}, \bibinfo {author}
  {\bibfnamefont {C.}~\bibnamefont {Shekhar}}, \bibinfo {author} {\bibfnamefont
  {Y.}~\bibnamefont {Sun}}, \bibinfo {author} {\bibfnamefont {U.}~\bibnamefont
  {Drechsler}}, \bibinfo {author} {\bibfnamefont {R.}~\bibnamefont {Zierold}},
  \bibinfo {author} {\bibfnamefont {C.}~\bibnamefont {Felser}}, \ and\ \bibinfo
  {author} {\bibfnamefont {B.}~\bibnamefont {Gotsmann}},\ }\bibfield  {title}
  {\enquote {\bibinfo {title} {Thermal and electrical signatures of a
  hydrodynamic electron fluid in tungsten diphosphide},}\ }\href {\doibase
  10.1038/s41467-018-06688-y} {\bibfield  {journal} {\bibinfo  {journal}
  {Nature Communications}\ }\textbf {\bibinfo {volume} {9}},\ \bibinfo {pages}
  {4093} (\bibinfo {year} {2018})},\ \Eprint {http://arxiv.org/abs/1706.05925}
  {arXiv:1706.05925 [cond-mat.str-el]} \BibitemShut {NoStop}%
\bibitem [{\citenamefont {Zaanen}(2019)}]{Zaanen:2018edk}%
  \BibitemOpen
  \bibfield  {author} {\bibinfo {author} {\bibfnamefont {Jan}\ \bibnamefont
  {Zaanen}},\ }\bibfield  {title} {\enquote {\bibinfo {title} {{Planckian
  dissipation, minimal viscosity and the transport in cuprate strange
  metals}},}\ }\href {\doibase 10.21468/SciPostPhys.6.5.061} {\bibfield
  {journal} {\bibinfo  {journal} {SciPost Phys.}\ }\textbf {\bibinfo {volume}
  {6}},\ \bibinfo {pages} {061} (\bibinfo {year} {2019})},\ \Eprint
  {http://arxiv.org/abs/1807.10951} {arXiv:1807.10951 [cond-mat.str-el]}
  \BibitemShut {NoStop}%
\bibitem [{\citenamefont {Sachdev}(2011{\natexlab{a}})}]{sachdevbook}%
  \BibitemOpen
  \bibfield  {author} {\bibinfo {author} {\bibfnamefont {Subir}\ \bibnamefont
  {Sachdev}},\ }\href {\doibase 10.1017/CBO9780511973765} {\emph {\bibinfo
  {title} {{Quantum phase transitions}}}}\ (\bibinfo  {publisher} {Cambridge
  University Press},\ \bibinfo {address} {Cambridge, UK},\ \bibinfo {year}
  {2011})\BibitemShut {NoStop}%
\bibitem [{\citenamefont {G\"otze}\ and\ \citenamefont
  {W\"olfle}(1972)}]{PhysRevB.6.1226}%
  \BibitemOpen
  \bibfield  {author} {\bibinfo {author} {\bibfnamefont {W.}~\bibnamefont
  {G\"otze}}\ and\ \bibinfo {author} {\bibfnamefont {P.}~\bibnamefont
  {W\"olfle}},\ }\bibfield  {title} {\enquote {\bibinfo {title} {Homogeneous
  dynamical conductivity of simple metals},}\ }\href {\doibase
  10.1103/PhysRevB.6.1226} {\bibfield  {journal} {\bibinfo  {journal} {Phys.
  Rev. B}\ }\textbf {\bibinfo {volume} {6}},\ \bibinfo {pages} {1226--1238}
  (\bibinfo {year} {1972})}\BibitemShut {NoStop}%
\bibitem [{\citenamefont {Rosch}\ and\ \citenamefont
  {Andrei}(2000)}]{PhysRevLett.85.1092}%
  \BibitemOpen
  \bibfield  {author} {\bibinfo {author} {\bibfnamefont {A.}~\bibnamefont
  {Rosch}}\ and\ \bibinfo {author} {\bibfnamefont {N.}~\bibnamefont {Andrei}},\
  }\bibfield  {title} {\enquote {\bibinfo {title} {Conductivity of a clean
  one-dimensional wire},}\ }\href {\doibase 10.1103/PhysRevLett.85.1092}
  {\bibfield  {journal} {\bibinfo  {journal} {Phys. Rev. Lett.}\ }\textbf
  {\bibinfo {volume} {85}},\ \bibinfo {pages} {1092--1095} (\bibinfo {year}
  {2000})}\BibitemShut {NoStop}%
\bibitem [{\citenamefont {Hartnoll}\ and\ \citenamefont
  {Hofman}(2012)}]{Hartnoll:2012rj}%
  \BibitemOpen
  \bibfield  {author} {\bibinfo {author} {\bibfnamefont {Sean~A.}\ \bibnamefont
  {Hartnoll}}\ and\ \bibinfo {author} {\bibfnamefont {Diego~M.}\ \bibnamefont
  {Hofman}},\ }\bibfield  {title} {\enquote {\bibinfo {title} {{Locally
  Critical Resistivities from Umklapp Scattering}},}\ }\href {\doibase
  10.1103/PhysRevLett.108.241601} {\bibfield  {journal} {\bibinfo  {journal}
  {Phys. Rev. Lett.}\ }\textbf {\bibinfo {volume} {108}},\ \bibinfo {pages}
  {241601} (\bibinfo {year} {2012})},\ \Eprint {http://arxiv.org/abs/1201.3917}
  {arXiv:1201.3917 [hep-th]} \BibitemShut {NoStop}%
\bibitem [{\citenamefont {Hartnoll}\ \emph {et~al.}(2014)\citenamefont
  {Hartnoll}, \citenamefont {Mahajan}, \citenamefont {Punk},\ and\
  \citenamefont {Sachdev}}]{Hartnoll:2014gba}%
  \BibitemOpen
  \bibfield  {author} {\bibinfo {author} {\bibfnamefont {Sean~A.}\ \bibnamefont
  {Hartnoll}}, \bibinfo {author} {\bibfnamefont {Raghu}\ \bibnamefont
  {Mahajan}}, \bibinfo {author} {\bibfnamefont {Matthias}\ \bibnamefont
  {Punk}}, \ and\ \bibinfo {author} {\bibfnamefont {Subir}\ \bibnamefont
  {Sachdev}},\ }\bibfield  {title} {\enquote {\bibinfo {title} {{Transport near
  the Ising-nematic quantum critical point of metals in two dimensions}},}\
  }\href {\doibase 10.1103/PhysRevB.89.155130} {\bibfield  {journal} {\bibinfo
  {journal} {Phys. Rev.}\ }\textbf {\bibinfo {volume} {B89}},\ \bibinfo {pages}
  {155130} (\bibinfo {year} {2014})},\ \Eprint {http://arxiv.org/abs/1401.7012}
  {arXiv:1401.7012 [cond-mat.str-el]} \BibitemShut {NoStop}%
\bibitem [{\citenamefont {Patel}\ and\ \citenamefont
  {Sachdev}(2014)}]{Patel:2014jfa}%
  \BibitemOpen
  \bibfield  {author} {\bibinfo {author} {\bibfnamefont {Aavishkar~A.}\
  \bibnamefont {Patel}}\ and\ \bibinfo {author} {\bibfnamefont {Subir}\
  \bibnamefont {Sachdev}},\ }\bibfield  {title} {\enquote {\bibinfo {title}
  {{DC resistivity at the onset of spin density wave order in two-dimensional
  metals}},}\ }\href {\doibase 10.1103/PhysRevB.90.165146} {\bibfield
  {journal} {\bibinfo  {journal} {Phys. Rev.}\ }\textbf {\bibinfo {volume}
  {B90}},\ \bibinfo {pages} {165146} (\bibinfo {year} {2014})},\ \Eprint
  {http://arxiv.org/abs/1408.6549} {arXiv:1408.6549 [cond-mat.str-el]}
  \BibitemShut {NoStop}%
\bibitem [{\citenamefont {Sachdev}(2011{\natexlab{b}})}]{inbooksachdev}%
  \BibitemOpen
  \bibfield  {author} {\bibinfo {author} {\bibfnamefont {Subir}\ \bibnamefont
  {Sachdev}},\ }\enquote {\bibinfo {title} {{Quantum phase transitions}},}\ \
  (\bibinfo  {publisher} {Cambridge University Press},\ \bibinfo {address}
  {Cambridge, UK},\ \bibinfo {year} {2011})\ Chap.~\bibinfo {chapter}
  {18}\BibitemShut {NoStop}%
\bibitem [{\citenamefont {Davison}\ \emph {et~al.}(2015)\citenamefont
  {Davison}, \citenamefont {Gout\'eraux},\ and\ \citenamefont
  {Hartnoll}}]{Davison:2015taa}%
  \BibitemOpen
  \bibfield  {author} {\bibinfo {author} {\bibfnamefont {Richard~A.}\
  \bibnamefont {Davison}}, \bibinfo {author} {\bibfnamefont {Blaise}\
  \bibnamefont {Gout\'eraux}}, \ and\ \bibinfo {author} {\bibfnamefont
  {Sean~A.}\ \bibnamefont {Hartnoll}},\ }\bibfield  {title} {\enquote {\bibinfo
  {title} {{Incoherent transport in clean quantum critical metals}},}\ }\href
  {\doibase 10.1007/JHEP10(2015)112} {\bibfield  {journal} {\bibinfo  {journal}
  {JHEP}\ }\textbf {\bibinfo {volume} {10}},\ \bibinfo {pages} {112} (\bibinfo
  {year} {2015})},\ \Eprint {http://arxiv.org/abs/1507.07137} {arXiv:1507.07137
  [hep-th]} \BibitemShut {NoStop}%
\bibitem [{\citenamefont
  {Gout\'eraux}(2014{\natexlab{a}})}]{Gouteraux:2014hca}%
  \BibitemOpen
  \bibfield  {author} {\bibinfo {author} {\bibfnamefont {B.}~\bibnamefont
  {Gout\'eraux}},\ }\bibfield  {title} {\enquote {\bibinfo {title} {{Charge
  transport in holography with momentum dissipation}},}\ }\href {\doibase
  10.1007/JHEP04(2014)181} {\bibfield  {journal} {\bibinfo  {journal} {JHEP}\
  }\textbf {\bibinfo {volume} {1404}},\ \bibinfo {pages} {181} (\bibinfo {year}
  {2014}{\natexlab{a}})},\ \Eprint {http://arxiv.org/abs/1401.5436}
  {arXiv:1401.5436 [hep-th]} \BibitemShut {NoStop}%
\bibitem [{\citenamefont {Davison}\ and\ \citenamefont
  {Gout\'eraux}(2015{\natexlab{a}})}]{Davison:2015bea}%
  \BibitemOpen
  \bibfield  {author} {\bibinfo {author} {\bibfnamefont {Richard~A.}\
  \bibnamefont {Davison}}\ and\ \bibinfo {author} {\bibfnamefont {Blaise}\
  \bibnamefont {Gout\'eraux}},\ }\bibfield  {title} {\enquote {\bibinfo {title}
  {{Dissecting holographic conductivities}},}\ }\href {\doibase
  10.1007/JHEP09(2015)090} {\bibfield  {journal} {\bibinfo  {journal} {JHEP}\
  }\textbf {\bibinfo {volume} {09}},\ \bibinfo {pages} {090} (\bibinfo {year}
  {2015}{\natexlab{a}})},\ \Eprint {http://arxiv.org/abs/1505.05092}
  {arXiv:1505.05092 [hep-th]} \BibitemShut {NoStop}%
\bibitem [{\citenamefont {Blake}\ \emph {et~al.}(2017)\citenamefont {Blake},
  \citenamefont {Davison},\ and\ \citenamefont {Sachdev}}]{Blake:2017qgd}%
  \BibitemOpen
  \bibfield  {author} {\bibinfo {author} {\bibfnamefont {Mike}\ \bibnamefont
  {Blake}}, \bibinfo {author} {\bibfnamefont {Richard~A.}\ \bibnamefont
  {Davison}}, \ and\ \bibinfo {author} {\bibfnamefont {Subir}\ \bibnamefont
  {Sachdev}},\ }\bibfield  {title} {\enquote {\bibinfo {title} {{Thermal
  diffusivity and chaos in metals without quasiparticles}},}\ }\href {\doibase
  10.1103/PhysRevD.96.106008} {\bibfield  {journal} {\bibinfo  {journal} {Phys.
  Rev.}\ }\textbf {\bibinfo {volume} {D96}},\ \bibinfo {pages} {106008}
  (\bibinfo {year} {2017})},\ \Eprint {http://arxiv.org/abs/1705.07896}
  {arXiv:1705.07896 [hep-th]} \BibitemShut {NoStop}%
\bibitem [{\citenamefont {Rosch}(1999)}]{PhysRevLett.82.4280}%
  \BibitemOpen
  \bibfield  {author} {\bibinfo {author} {\bibfnamefont {A.}~\bibnamefont
  {Rosch}},\ }\bibfield  {title} {\enquote {\bibinfo {title} {Interplay of
  disorder and spin fluctuations in the resistivity near a quantum critical
  point},}\ }\href {\doibase 10.1103/PhysRevLett.82.4280} {\bibfield  {journal}
  {\bibinfo  {journal} {Phys. Rev. Lett.}\ }\textbf {\bibinfo {volume} {82}},\
  \bibinfo {pages} {4280--4283} (\bibinfo {year} {1999})}\BibitemShut {NoStop}%
\bibitem [{\citenamefont {Sirker}\ \emph {et~al.}(2009)\citenamefont {Sirker},
  \citenamefont {Pereira},\ and\ \citenamefont
  {Affleck}}]{PhysRevLett.103.216602}%
  \BibitemOpen
  \bibfield  {author} {\bibinfo {author} {\bibfnamefont {J.}~\bibnamefont
  {Sirker}}, \bibinfo {author} {\bibfnamefont {R.~G.}\ \bibnamefont {Pereira}},
  \ and\ \bibinfo {author} {\bibfnamefont {I.}~\bibnamefont {Affleck}},\
  }\bibfield  {title} {\enquote {\bibinfo {title} {Diffusion and ballistic
  transport in one-dimensional quantum systems},}\ }\href {\doibase
  10.1103/PhysRevLett.103.216602} {\bibfield  {journal} {\bibinfo  {journal}
  {Phys. Rev. Lett.}\ }\textbf {\bibinfo {volume} {103}},\ \bibinfo {pages}
  {216602} (\bibinfo {year} {2009})}\BibitemShut {NoStop}%
\bibitem [{\citenamefont {Hartnoll}\ \emph {et~al.}(2011)\citenamefont
  {Hartnoll}, \citenamefont {Hofman}, \citenamefont {Metlitski},\ and\
  \citenamefont {Sachdev}}]{Hartnoll:2011ic}%
  \BibitemOpen
  \bibfield  {author} {\bibinfo {author} {\bibfnamefont {Sean~A.}\ \bibnamefont
  {Hartnoll}}, \bibinfo {author} {\bibfnamefont {Diego~M.}\ \bibnamefont
  {Hofman}}, \bibinfo {author} {\bibfnamefont {Max~A.}\ \bibnamefont
  {Metlitski}}, \ and\ \bibinfo {author} {\bibfnamefont {Subir}\ \bibnamefont
  {Sachdev}},\ }\bibfield  {title} {\enquote {\bibinfo {title} {{Quantum
  critical response at the onset of spin density wave order in two-dimensional
  metals}},}\ }\href {\doibase 10.1103/PhysRevB.84.125115} {\bibfield
  {journal} {\bibinfo  {journal} {Phys. Rev.}\ }\textbf {\bibinfo {volume}
  {B84}},\ \bibinfo {pages} {125115} (\bibinfo {year} {2011})},\ \Eprint
  {http://arxiv.org/abs/1106.0001} {arXiv:1106.0001 [cond-mat.str-el]}
  \BibitemShut {NoStop}%
\bibitem [{\citenamefont {Charmousis}\ \emph {et~al.}(2010)\citenamefont
  {Charmousis}, \citenamefont {Gout\'eraux}, \citenamefont {Kim}, \citenamefont
  {Kiritsis},\ and\ \citenamefont {Meyer}}]{Charmousis:2010zz}%
  \BibitemOpen
  \bibfield  {author} {\bibinfo {author} {\bibfnamefont {Christos}\
  \bibnamefont {Charmousis}}, \bibinfo {author} {\bibfnamefont {Blaise}\
  \bibnamefont {Gout\'eraux}}, \bibinfo {author} {\bibfnamefont {Bom~Soo}\
  \bibnamefont {Kim}}, \bibinfo {author} {\bibfnamefont {Elias}\ \bibnamefont
  {Kiritsis}}, \ and\ \bibinfo {author} {\bibfnamefont {Rene}\ \bibnamefont
  {Meyer}},\ }\bibfield  {title} {\enquote {\bibinfo {title} {{Effective
  Holographic Theories for low-temperature condensed matter systems}},}\ }\href
  {\doibase 10.1007/JHEP11(2010)151} {\bibfield  {journal} {\bibinfo  {journal}
  {JHEP}\ }\textbf {\bibinfo {volume} {1011}},\ \bibinfo {pages} {151}
  (\bibinfo {year} {2010})},\ \Eprint {http://arxiv.org/abs/1005.4690}
  {arXiv:1005.4690 [hep-th]} \BibitemShut {NoStop}%
\bibitem [{\citenamefont {Gout\'eraux}\ and\ \citenamefont
  {Kiritsis}(2011)}]{Gouteraux:2011ce}%
  \BibitemOpen
  \bibfield  {author} {\bibinfo {author} {\bibfnamefont {B.}~\bibnamefont
  {Gout\'eraux}}\ and\ \bibinfo {author} {\bibfnamefont {E.}~\bibnamefont
  {Kiritsis}},\ }\bibfield  {title} {\enquote {\bibinfo {title} {{Generalized
  Holographic Quantum Criticality at Finite Density}},}\ }\href {\doibase
  10.1007/JHEP12(2011)036} {\bibfield  {journal} {\bibinfo  {journal} {JHEP}\
  }\textbf {\bibinfo {volume} {1112}},\ \bibinfo {pages} {036} (\bibinfo {year}
  {2011})},\ \Eprint {http://arxiv.org/abs/1107.2116} {arXiv:1107.2116
  [hep-th]} \BibitemShut {NoStop}%
\bibitem [{\citenamefont {Hartman}\ \emph {et~al.}(2017)\citenamefont
  {Hartman}, \citenamefont {Hartnoll},\ and\ \citenamefont
  {Mahajan}}]{Hartman:2017hhp}%
  \BibitemOpen
  \bibfield  {author} {\bibinfo {author} {\bibfnamefont {Thomas}\ \bibnamefont
  {Hartman}}, \bibinfo {author} {\bibfnamefont {Sean~A.}\ \bibnamefont
  {Hartnoll}}, \ and\ \bibinfo {author} {\bibfnamefont {Raghu}\ \bibnamefont
  {Mahajan}},\ }\bibfield  {title} {\enquote {\bibinfo {title} {{Upper Bound on
  Diffusivity}},}\ }\href {\doibase 10.1103/PhysRevLett.119.141601} {\bibfield
  {journal} {\bibinfo  {journal} {Phys. Rev. Lett.}\ }\textbf {\bibinfo
  {volume} {119}},\ \bibinfo {pages} {141601} (\bibinfo {year} {2017})},\
  \Eprint {http://arxiv.org/abs/1706.00019} {arXiv:1706.00019 [hep-th]}
  \BibitemShut {NoStop}%
\bibitem [{\citenamefont {Lucas}(2017)}]{Lucas:2017ibu}%
  \BibitemOpen
  \bibfield  {author} {\bibinfo {author} {\bibfnamefont {Andrew}\ \bibnamefont
  {Lucas}},\ }\bibfield  {title} {\enquote {\bibinfo {title} {{Constraints on
  hydrodynamics from many-body quantum chaos}},}\ }\href@noop {} {\  (\bibinfo
  {year} {2017})},\ \Eprint {http://arxiv.org/abs/1710.01005} {arXiv:1710.01005
  [hep-th]} \BibitemShut {NoStop}%
\bibitem [{\citenamefont {Blake}(2016{\natexlab{a}})}]{Blake:2016wvh}%
  \BibitemOpen
  \bibfield  {author} {\bibinfo {author} {\bibfnamefont {Mike}\ \bibnamefont
  {Blake}},\ }\bibfield  {title} {\enquote {\bibinfo {title} {{Universal Charge
  Diffusion and the Butterfly Effect in Holographic Theories}},}\ }\href
  {\doibase 10.1103/PhysRevLett.117.091601} {\bibfield  {journal} {\bibinfo
  {journal} {Phys. Rev. Lett.}\ }\textbf {\bibinfo {volume} {117}},\ \bibinfo
  {pages} {091601} (\bibinfo {year} {2016}{\natexlab{a}})},\ \Eprint
  {http://arxiv.org/abs/1603.08510} {arXiv:1603.08510 [hep-th]} \BibitemShut
  {NoStop}%
\bibitem [{\citenamefont {Blake}(2016{\natexlab{b}})}]{Blake:2016sud}%
  \BibitemOpen
  \bibfield  {author} {\bibinfo {author} {\bibfnamefont {Mike}\ \bibnamefont
  {Blake}},\ }\bibfield  {title} {\enquote {\bibinfo {title} {{Universal
  Diffusion in Incoherent Black Holes}},}\ }\href {\doibase
  10.1103/PhysRevD.94.086014} {\bibfield  {journal} {\bibinfo  {journal} {Phys.
  Rev.}\ }\textbf {\bibinfo {volume} {D94}},\ \bibinfo {pages} {086014}
  (\bibinfo {year} {2016}{\natexlab{b}})},\ \Eprint
  {http://arxiv.org/abs/1604.01754} {arXiv:1604.01754 [hep-th]} \BibitemShut
  {NoStop}%
\bibitem [{\citenamefont {Gu}\ \emph {et~al.}(2017)\citenamefont {Gu},
  \citenamefont {Qi},\ and\ \citenamefont {Stanford}}]{Gu:2016oyy}%
  \BibitemOpen
  \bibfield  {author} {\bibinfo {author} {\bibfnamefont {Yingfei}\ \bibnamefont
  {Gu}}, \bibinfo {author} {\bibfnamefont {Xiao-Liang}\ \bibnamefont {Qi}}, \
  and\ \bibinfo {author} {\bibfnamefont {Douglas}\ \bibnamefont {Stanford}},\
  }\bibfield  {title} {\enquote {\bibinfo {title} {{Local criticality,
  diffusion and chaos in generalized Sachdev-Ye-Kitaev models}},}\ }\href
  {\doibase 10.1007/JHEP05(2017)125} {\bibfield  {journal} {\bibinfo  {journal}
  {JHEP}\ }\textbf {\bibinfo {volume} {05}},\ \bibinfo {pages} {125} (\bibinfo
  {year} {2017})},\ \Eprint {http://arxiv.org/abs/1609.07832} {arXiv:1609.07832
  [hep-th]} \BibitemShut {NoStop}%
\bibitem [{\citenamefont {Blake}\ and\ \citenamefont
  {Donos}(2017)}]{Blake:2016jnn}%
  \BibitemOpen
  \bibfield  {author} {\bibinfo {author} {\bibfnamefont {Mike}\ \bibnamefont
  {Blake}}\ and\ \bibinfo {author} {\bibfnamefont {Aristomenis}\ \bibnamefont
  {Donos}},\ }\bibfield  {title} {\enquote {\bibinfo {title} {{Diffusion and
  Chaos from near AdS$_2$ horizons}},}\ }\href {\doibase
  10.1007/JHEP02(2017)013} {\bibfield  {journal} {\bibinfo  {journal} {JHEP}\
  }\textbf {\bibinfo {volume} {02}},\ \bibinfo {pages} {013} (\bibinfo {year}
  {2017})},\ \Eprint {http://arxiv.org/abs/1611.09380} {arXiv:1611.09380
  [hep-th]} \BibitemShut {NoStop}%
\bibitem [{\citenamefont {Davison}\ \emph {et~al.}(2017)\citenamefont
  {Davison}, \citenamefont {Fu}, \citenamefont {Georges}, \citenamefont {Gu},
  \citenamefont {Jensen},\ and\ \citenamefont {Sachdev}}]{Davison:2016ngz}%
  \BibitemOpen
  \bibfield  {author} {\bibinfo {author} {\bibfnamefont {Richard~A.}\
  \bibnamefont {Davison}}, \bibinfo {author} {\bibfnamefont {Wenbo}\
  \bibnamefont {Fu}}, \bibinfo {author} {\bibfnamefont {Antoine}\ \bibnamefont
  {Georges}}, \bibinfo {author} {\bibfnamefont {Yingfei}\ \bibnamefont {Gu}},
  \bibinfo {author} {\bibfnamefont {Kristan}\ \bibnamefont {Jensen}}, \ and\
  \bibinfo {author} {\bibfnamefont {Subir}\ \bibnamefont {Sachdev}},\
  }\bibfield  {title} {\enquote {\bibinfo {title} {{Thermoelectric transport in
  disordered metals without quasiparticles: The Sachdev-Ye-Kitaev models and
  holography}},}\ }\href {\doibase 10.1103/PhysRevB.95.155131} {\bibfield
  {journal} {\bibinfo  {journal} {Phys. Rev.}\ }\textbf {\bibinfo {volume}
  {B95}},\ \bibinfo {pages} {155131} (\bibinfo {year} {2017})},\ \Eprint
  {http://arxiv.org/abs/1612.00849} {arXiv:1612.00849 [cond-mat.str-el]}
  \BibitemShut {NoStop}%
\bibitem [{\citenamefont {Patel}\ and\ \citenamefont
  {Sachdev}(2017)}]{Patel:2016wdy}%
  \BibitemOpen
  \bibfield  {author} {\bibinfo {author} {\bibfnamefont {Aavishkar~A.}\
  \bibnamefont {Patel}}\ and\ \bibinfo {author} {\bibfnamefont {Subir}\
  \bibnamefont {Sachdev}},\ }\bibfield  {title} {\enquote {\bibinfo {title}
  {{Quantum chaos on a critical Fermi surface}},}\ }\href {\doibase
  10.1073/pnas.1618185114} {\bibfield  {journal} {\bibinfo  {journal} {Proc.
  Nat. Acad. Sci.}\ }\textbf {\bibinfo {volume} {114}},\ \bibinfo {pages}
  {1844--1849} (\bibinfo {year} {2017})},\ \Eprint
  {http://arxiv.org/abs/1611.00003} {arXiv:1611.00003 [cond-mat.str-el]}
  \BibitemShut {NoStop}%
\bibitem [{\citenamefont {Werman}\ \emph {et~al.}(2017)\citenamefont {Werman},
  \citenamefont {Kivelson},\ and\ \citenamefont {Berg}}]{Werman:2017abn}%
  \BibitemOpen
  \bibfield  {author} {\bibinfo {author} {\bibfnamefont {Yochai}\ \bibnamefont
  {Werman}}, \bibinfo {author} {\bibfnamefont {Steven~A.}\ \bibnamefont
  {Kivelson}}, \ and\ \bibinfo {author} {\bibfnamefont {Erez}\ \bibnamefont
  {Berg}},\ }\bibfield  {title} {\enquote {\bibinfo {title} {{Quantum chaos in
  an electron-phonon bad metal}},}\ }\href@noop {} {\  (\bibinfo {year}
  {2017})},\ \Eprint {http://arxiv.org/abs/1705.07895} {arXiv:1705.07895
  [cond-mat.str-el]} \BibitemShut {NoStop}%
\bibitem [{\citenamefont {Blake}\ \emph
  {et~al.}(2018{\natexlab{a}})\citenamefont {Blake}, \citenamefont {Lee},\ and\
  \citenamefont {Liu}}]{Blake:2017ris}%
  \BibitemOpen
  \bibfield  {author} {\bibinfo {author} {\bibfnamefont {Mike}\ \bibnamefont
  {Blake}}, \bibinfo {author} {\bibfnamefont {Hyunseok}\ \bibnamefont {Lee}}, \
  and\ \bibinfo {author} {\bibfnamefont {Hong}\ \bibnamefont {Liu}},\
  }\bibfield  {title} {\enquote {\bibinfo {title} {{A quantum hydrodynamical
  description for scrambling and many-body chaos}},}\ }\href {\doibase
  10.1007/JHEP10(2018)127} {\bibfield  {journal} {\bibinfo  {journal} {JHEP}\
  }\textbf {\bibinfo {volume} {10}},\ \bibinfo {pages} {127} (\bibinfo {year}
  {2018}{\natexlab{a}})},\ \Eprint {http://arxiv.org/abs/1801.00010}
  {arXiv:1801.00010 [hep-th]} \BibitemShut {NoStop}%
\bibitem [{\citenamefont {Blake}\ \emph
  {et~al.}(2018{\natexlab{b}})\citenamefont {Blake}, \citenamefont {Davison},
  \citenamefont {Grozdanov},\ and\ \citenamefont {Liu}}]{Blake:2018leo}%
  \BibitemOpen
  \bibfield  {author} {\bibinfo {author} {\bibfnamefont {Mike}\ \bibnamefont
  {Blake}}, \bibinfo {author} {\bibfnamefont {Richard~A.}\ \bibnamefont
  {Davison}}, \bibinfo {author} {\bibfnamefont {Sa\v{s}o}\ \bibnamefont
  {Grozdanov}}, \ and\ \bibinfo {author} {\bibfnamefont {Hong}\ \bibnamefont
  {Liu}},\ }\bibfield  {title} {\enquote {\bibinfo {title} {{Many-body chaos
  and energy dynamics in holography}},}\ }\href {\doibase
  10.1007/JHEP10(2018)035} {\bibfield  {journal} {\bibinfo  {journal} {JHEP}\
  }\textbf {\bibinfo {volume} {10}},\ \bibinfo {pages} {035} (\bibinfo {year}
  {2018}{\natexlab{b}})},\ \Eprint {http://arxiv.org/abs/1809.01169}
  {arXiv:1809.01169 [hep-th]} \BibitemShut {NoStop}%
\bibitem [{\citenamefont
  {Gout\'eraux}(2014{\natexlab{b}})}]{Gouteraux:2013oca}%
  \BibitemOpen
  \bibfield  {author} {\bibinfo {author} {\bibfnamefont {B.}~\bibnamefont
  {Gout\'eraux}},\ }\bibfield  {title} {\enquote {\bibinfo {title} {{Universal
  scaling properties of extremal cohesive holographic phases}},}\ }\href
  {\doibase 10.1007/JHEP01(2014)080} {\bibfield  {journal} {\bibinfo  {journal}
  {JHEP}\ }\textbf {\bibinfo {volume} {1401}},\ \bibinfo {pages} {080}
  (\bibinfo {year} {2014}{\natexlab{b}})},\ \Eprint
  {http://arxiv.org/abs/1308.2084} {arXiv:1308.2084 [hep-th]} \BibitemShut
  {NoStop}%
\bibitem [{\citenamefont {Karch}(2014)}]{Karch:2014mba}%
  \BibitemOpen
  \bibfield  {author} {\bibinfo {author} {\bibfnamefont {Andreas}\ \bibnamefont
  {Karch}},\ }\bibfield  {title} {\enquote {\bibinfo {title} {{Conductivities
  for Hyperscaling Violating Geometries}},}\ }\href {\doibase
  10.1007/JHEP06(2014)140} {\bibfield  {journal} {\bibinfo  {journal} {JHEP}\
  }\textbf {\bibinfo {volume} {1406}},\ \bibinfo {pages} {140} (\bibinfo {year}
  {2014})},\ \Eprint {http://arxiv.org/abs/1405.2926} {arXiv:1405.2926
  [hep-th]} \BibitemShut {NoStop}%
\bibitem [{\citenamefont {Davison}\ \emph {et~al.}(2018)\citenamefont
  {Davison}, \citenamefont {Gentle},\ and\ \citenamefont
  {Gout\'eraux}}]{Davison:2018}%
  \BibitemOpen
  \bibfield  {author} {\bibinfo {author} {\bibfnamefont {Richard~A.}\
  \bibnamefont {Davison}}, \bibinfo {author} {\bibfnamefont {Simon~A.}\
  \bibnamefont {Gentle}}, \ and\ \bibinfo {author} {\bibfnamefont {Blaise}\
  \bibnamefont {Gout\'eraux}},\ }\bibfield  {title} {\enquote {\bibinfo {title}
  {{Impact of irrelevant deformations on thermodynamics and transport in
  holographic quantum critical states}},}\ }\href@noop {} {\  (\bibinfo {year}
  {2018})},\ \Eprint {http://arxiv.org/abs/1812.11060} {arXiv:1812.11060
  [hep-th]} \BibitemShut {NoStop}%
\bibitem [{\citenamefont {Hartnoll}\ and\ \citenamefont
  {Karch}(2015)}]{Hartnoll:2015sea}%
  \BibitemOpen
  \bibfield  {author} {\bibinfo {author} {\bibfnamefont {Sean~A.}\ \bibnamefont
  {Hartnoll}}\ and\ \bibinfo {author} {\bibfnamefont {Andreas}\ \bibnamefont
  {Karch}},\ }\bibfield  {title} {\enquote {\bibinfo {title} {{Scaling theory
  of the cuprate strange metals}},}\ }\href {\doibase
  10.1103/PhysRevB.91.155126} {\bibfield  {journal} {\bibinfo  {journal} {Phys.
  Rev.}\ }\textbf {\bibinfo {volume} {B91}},\ \bibinfo {pages} {155126}
  (\bibinfo {year} {2015})},\ \Eprint {http://arxiv.org/abs/1501.03165}
  {arXiv:1501.03165 [cond-mat.str-el]} \BibitemShut {NoStop}%
\bibitem [{\citenamefont {{Phillips}}\ and\ \citenamefont
  {{Chamon}}(2005)}]{2005PhRvL..95j7002P}%
  \BibitemOpen
  \bibfield  {author} {\bibinfo {author} {\bibfnamefont {P.}~\bibnamefont
  {{Phillips}}}\ and\ \bibinfo {author} {\bibfnamefont {C.}~\bibnamefont
  {{Chamon}}},\ }\bibfield  {title} {\enquote {\bibinfo {title} {{Breakdown of
  One-Parameter Scaling in Quantum Critical Scenarios for High-Temperature
  Copper-Oxide Superconductors}},}\ }\href {\doibase
  10.1103/PhysRevLett.95.107002} {\bibfield  {journal} {\bibinfo  {journal}
  {Physical Review Letters}\ }\textbf {\bibinfo {volume} {95}},\ \bibinfo {eid}
  {107002} (\bibinfo {year} {2005})},\ \Eprint
  {http://arxiv.org/abs/cond-mat/0412179} {cond-mat/0412179} \BibitemShut
  {NoStop}%
\bibitem [{\citenamefont {Huijse}\ \emph {et~al.}(2012)\citenamefont {Huijse},
  \citenamefont {Sachdev},\ and\ \citenamefont {Swingle}}]{Huijse:2011ef}%
  \BibitemOpen
  \bibfield  {author} {\bibinfo {author} {\bibfnamefont {Liza}\ \bibnamefont
  {Huijse}}, \bibinfo {author} {\bibfnamefont {Subir}\ \bibnamefont {Sachdev}},
  \ and\ \bibinfo {author} {\bibfnamefont {Brian}\ \bibnamefont {Swingle}},\
  }\bibfield  {title} {\enquote {\bibinfo {title} {{Hidden Fermi surfaces in
  compressible states of gauge-gravity duality}},}\ }\href {\doibase
  10.1103/PhysRevB.85.035121} {\bibfield  {journal} {\bibinfo  {journal}
  {Phys.Rev.}\ }\textbf {\bibinfo {volume} {B85}},\ \bibinfo {pages} {035121}
  (\bibinfo {year} {2012})},\ \Eprint {http://arxiv.org/abs/1112.0573}
  {arXiv:1112.0573 [cond-mat.str-el]} \BibitemShut {NoStop}%
\bibitem [{\citenamefont {Kanitscheider}\ and\ \citenamefont
  {Skenderis}(2009)}]{Kanitscheider:2009as}%
  \BibitemOpen
  \bibfield  {author} {\bibinfo {author} {\bibfnamefont {Ingmar}\ \bibnamefont
  {Kanitscheider}}\ and\ \bibinfo {author} {\bibfnamefont {Kostas}\
  \bibnamefont {Skenderis}},\ }\bibfield  {title} {\enquote {\bibinfo {title}
  {{Universal hydrodynamics of non-conformal branes}},}\ }\href {\doibase
  10.1088/1126-6708/2009/04/062} {\bibfield  {journal} {\bibinfo  {journal}
  {JHEP}\ }\textbf {\bibinfo {volume} {04}},\ \bibinfo {pages} {062} (\bibinfo
  {year} {2009})},\ \Eprint {http://arxiv.org/abs/0901.1487} {arXiv:0901.1487
  [hep-th]} \BibitemShut {NoStop}%
\bibitem [{\citenamefont {Gout\'eraux}\ \emph {et~al.}(2012)\citenamefont
  {Gout\'eraux}, \citenamefont {Smolic}, \citenamefont {Smolic}, \citenamefont
  {Skenderis},\ and\ \citenamefont {Taylor}}]{Gouteraux:2011qh}%
  \BibitemOpen
  \bibfield  {author} {\bibinfo {author} {\bibfnamefont {Blaise}\ \bibnamefont
  {Gout\'eraux}}, \bibinfo {author} {\bibfnamefont {Jelena}\ \bibnamefont
  {Smolic}}, \bibinfo {author} {\bibfnamefont {Milena}\ \bibnamefont {Smolic}},
  \bibinfo {author} {\bibfnamefont {Kostas}\ \bibnamefont {Skenderis}}, \ and\
  \bibinfo {author} {\bibfnamefont {Marika}\ \bibnamefont {Taylor}},\
  }\bibfield  {title} {\enquote {\bibinfo {title} {{Holography for
  Einstein-Maxwell-dilaton theories from generalized dimensional reduction}},}\
  }\href {\doibase 10.1007/JHEP01(2012)089} {\bibfield  {journal} {\bibinfo
  {journal} {JHEP}\ }\textbf {\bibinfo {volume} {01}},\ \bibinfo {pages} {089}
  (\bibinfo {year} {2012})},\ \Eprint {http://arxiv.org/abs/1110.2320}
  {arXiv:1110.2320 [hep-th]} \BibitemShut {NoStop}%
\bibitem [{\citenamefont {Hartnoll}\ and\ \citenamefont
  {Huijse}(2012)}]{Hartnoll:2011pp}%
  \BibitemOpen
  \bibfield  {author} {\bibinfo {author} {\bibfnamefont {Sean~A.}\ \bibnamefont
  {Hartnoll}}\ and\ \bibinfo {author} {\bibfnamefont {Liza}\ \bibnamefont
  {Huijse}},\ }\bibfield  {title} {\enquote {\bibinfo {title}
  {{Fractionalization of holographic Fermi surfaces}},}\ }\href {\doibase
  10.1088/0264-9381/29/19/194001} {\bibfield  {journal} {\bibinfo  {journal}
  {Class. Quant. Grav.}\ }\textbf {\bibinfo {volume} {29}},\ \bibinfo {pages}
  {194001} (\bibinfo {year} {2012})},\ \Eprint {http://arxiv.org/abs/1111.2606}
  {arXiv:1111.2606 [hep-th]} \BibitemShut {NoStop}%
\bibitem [{\citenamefont {Adam}\ \emph {et~al.}(2013)\citenamefont {Adam},
  \citenamefont {Crampton}, \citenamefont {Sonner},\ and\ \citenamefont
  {Withers}}]{Adam:2012mw}%
  \BibitemOpen
  \bibfield  {author} {\bibinfo {author} {\bibfnamefont {Alexander}\
  \bibnamefont {Adam}}, \bibinfo {author} {\bibfnamefont {Benedict}\
  \bibnamefont {Crampton}}, \bibinfo {author} {\bibfnamefont {Julian}\
  \bibnamefont {Sonner}}, \ and\ \bibinfo {author} {\bibfnamefont {Benjamin}\
  \bibnamefont {Withers}},\ }\bibfield  {title} {\enquote {\bibinfo {title}
  {{Bosonic Fractionalisation Transitions}},}\ }\href {\doibase
  10.1007/JHEP01(2013)127} {\bibfield  {journal} {\bibinfo  {journal} {JHEP}\
  }\textbf {\bibinfo {volume} {01}},\ \bibinfo {pages} {127} (\bibinfo {year}
  {2013})},\ \Eprint {http://arxiv.org/abs/1208.3199} {arXiv:1208.3199
  [hep-th]} \BibitemShut {NoStop}%
\bibitem [{\citenamefont {Gout\'eraux}\ and\ \citenamefont
  {Kiritsis}(2013)}]{Gouteraux:2012yr}%
  \BibitemOpen
  \bibfield  {author} {\bibinfo {author} {\bibfnamefont {B.}~\bibnamefont
  {Gout\'eraux}}\ and\ \bibinfo {author} {\bibfnamefont {E.}~\bibnamefont
  {Kiritsis}},\ }\bibfield  {title} {\enquote {\bibinfo {title} {{Quantum
  critical lines in holographic phases with (un)broken symmetry}},}\ }\href
  {\doibase 10.1007/JHEP04(2013)053} {\bibfield  {journal} {\bibinfo  {journal}
  {JHEP}\ }\textbf {\bibinfo {volume} {04}},\ \bibinfo {pages} {053} (\bibinfo
  {year} {2013})},\ \Eprint {http://arxiv.org/abs/1212.2625} {arXiv:1212.2625
  [hep-th]} \BibitemShut {NoStop}%
\bibitem [{\citenamefont {Policastro}\ \emph {et~al.}(2002)\citenamefont
  {Policastro}, \citenamefont {Son},\ and\ \citenamefont
  {Starinets}}]{Policastro:2002se}%
  \BibitemOpen
  \bibfield  {author} {\bibinfo {author} {\bibfnamefont {Giuseppe}\
  \bibnamefont {Policastro}}, \bibinfo {author} {\bibfnamefont {Dam~T.}\
  \bibnamefont {Son}}, \ and\ \bibinfo {author} {\bibfnamefont {Andrei~O.}\
  \bibnamefont {Starinets}},\ }\bibfield  {title} {\enquote {\bibinfo {title}
  {{From AdS / CFT correspondence to hydrodynamics}},}\ }\href {\doibase
  10.1088/1126-6708/2002/09/043} {\bibfield  {journal} {\bibinfo  {journal}
  {JHEP}\ }\textbf {\bibinfo {volume} {09}},\ \bibinfo {pages} {043} (\bibinfo
  {year} {2002})},\ \Eprint {http://arxiv.org/abs/hep-th/0205052}
  {arXiv:hep-th/0205052 [hep-th]} \BibitemShut {NoStop}%
\bibitem [{\citenamefont {Kovtun}\ and\ \citenamefont
  {Starinets}(2005)}]{Kovtun:2005ev}%
  \BibitemOpen
  \bibfield  {author} {\bibinfo {author} {\bibfnamefont {Pavel~K.}\
  \bibnamefont {Kovtun}}\ and\ \bibinfo {author} {\bibfnamefont {Andrei~O.}\
  \bibnamefont {Starinets}},\ }\bibfield  {title} {\enquote {\bibinfo {title}
  {{Quasinormal modes and holography}},}\ }\href {\doibase
  10.1103/PhysRevD.72.086009} {\bibfield  {journal} {\bibinfo  {journal} {Phys.
  Rev.}\ }\textbf {\bibinfo {volume} {D72}},\ \bibinfo {pages} {086009}
  (\bibinfo {year} {2005})},\ \Eprint {http://arxiv.org/abs/hep-th/0506184}
  {arXiv:hep-th/0506184 [hep-th]} \BibitemShut {NoStop}%
\bibitem [{\citenamefont {Edalati}\ \emph
  {et~al.}(2010{\natexlab{a}})\citenamefont {Edalati}, \citenamefont {Jottar},\
  and\ \citenamefont {Leigh}}]{Edalati:2010hk}%
  \BibitemOpen
  \bibfield  {author} {\bibinfo {author} {\bibfnamefont {Mohammad}\
  \bibnamefont {Edalati}}, \bibinfo {author} {\bibfnamefont {Juan~I.}\
  \bibnamefont {Jottar}}, \ and\ \bibinfo {author} {\bibfnamefont {Robert~G.}\
  \bibnamefont {Leigh}},\ }\bibfield  {title} {\enquote {\bibinfo {title}
  {{Shear Modes, Criticality and Extremal Black Holes}},}\ }\href {\doibase
  10.1007/JHEP04(2010)075} {\bibfield  {journal} {\bibinfo  {journal} {JHEP}\
  }\textbf {\bibinfo {volume} {04}},\ \bibinfo {pages} {075} (\bibinfo {year}
  {2010}{\natexlab{a}})},\ \Eprint {http://arxiv.org/abs/1001.0779}
  {arXiv:1001.0779 [hep-th]} \BibitemShut {NoStop}%
\bibitem [{\citenamefont {Edalati}\ \emph
  {et~al.}(2010{\natexlab{b}})\citenamefont {Edalati}, \citenamefont {Jottar},\
  and\ \citenamefont {Leigh}}]{Edalati:2010pn}%
  \BibitemOpen
  \bibfield  {author} {\bibinfo {author} {\bibfnamefont {Mohammad}\
  \bibnamefont {Edalati}}, \bibinfo {author} {\bibfnamefont {Juan~I.}\
  \bibnamefont {Jottar}}, \ and\ \bibinfo {author} {\bibfnamefont {Robert~G.}\
  \bibnamefont {Leigh}},\ }\bibfield  {title} {\enquote {\bibinfo {title}
  {{Holography and the sound of criticality}},}\ }\href {\doibase
  10.1007/JHEP10(2010)058} {\bibfield  {journal} {\bibinfo  {journal} {JHEP}\
  }\textbf {\bibinfo {volume} {10}},\ \bibinfo {pages} {058} (\bibinfo {year}
  {2010}{\natexlab{b}})},\ \Eprint {http://arxiv.org/abs/1005.4075}
  {arXiv:1005.4075 [hep-th]} \BibitemShut {NoStop}%
\bibitem [{\citenamefont {Buchel}\ \emph {et~al.}(2015)\citenamefont {Buchel},
  \citenamefont {Heller},\ and\ \citenamefont {Myers}}]{Buchel:2015saa}%
  \BibitemOpen
  \bibfield  {author} {\bibinfo {author} {\bibfnamefont {Alex}\ \bibnamefont
  {Buchel}}, \bibinfo {author} {\bibfnamefont {Michal~P.}\ \bibnamefont
  {Heller}}, \ and\ \bibinfo {author} {\bibfnamefont {Robert~C.}\ \bibnamefont
  {Myers}},\ }\bibfield  {title} {\enquote {\bibinfo {title} {{Equilibration
  rates in a strongly coupled nonconformal quark-gluon plasma}},}\ }\href
  {\doibase 10.1103/PhysRevLett.114.251601} {\bibfield  {journal} {\bibinfo
  {journal} {Phys. Rev. Lett.}\ }\textbf {\bibinfo {volume} {114}},\ \bibinfo
  {pages} {251601} (\bibinfo {year} {2015})},\ \Eprint
  {http://arxiv.org/abs/1503.07114} {arXiv:1503.07114 [hep-th]} \BibitemShut
  {NoStop}%
\bibitem [{\citenamefont {Janik}\ \emph {et~al.}(2015)\citenamefont {Janik},
  \citenamefont {Plewa}, \citenamefont {Soltanpanahi},\ and\ \citenamefont
  {Spalinski}}]{Janik:2015waa}%
  \BibitemOpen
  \bibfield  {author} {\bibinfo {author} {\bibfnamefont {Romuald~A.}\
  \bibnamefont {Janik}}, \bibinfo {author} {\bibfnamefont {Grzegorz}\
  \bibnamefont {Plewa}}, \bibinfo {author} {\bibfnamefont {Hesam}\ \bibnamefont
  {Soltanpanahi}}, \ and\ \bibinfo {author} {\bibfnamefont {Michal}\
  \bibnamefont {Spalinski}},\ }\bibfield  {title} {\enquote {\bibinfo {title}
  {{Linearized nonequilibrium dynamics in nonconformal plasma}},}\ }\href
  {\doibase 10.1103/PhysRevD.91.126013} {\bibfield  {journal} {\bibinfo
  {journal} {Phys. Rev.}\ }\textbf {\bibinfo {volume} {D91}},\ \bibinfo {pages}
  {126013} (\bibinfo {year} {2015})},\ \Eprint
  {http://arxiv.org/abs/1503.07149} {arXiv:1503.07149 [hep-th]} \BibitemShut
  {NoStop}%
\bibitem [{\citenamefont {Buchel}\ and\ \citenamefont
  {Day}(2015)}]{Buchel:2015ofa}%
  \BibitemOpen
  \bibfield  {author} {\bibinfo {author} {\bibfnamefont {Alex}\ \bibnamefont
  {Buchel}}\ and\ \bibinfo {author} {\bibfnamefont {Andrew}\ \bibnamefont
  {Day}},\ }\bibfield  {title} {\enquote {\bibinfo {title} {{Universal
  relaxation in quark-gluon plasma at strong coupling}},}\ }\href {\doibase
  10.1103/PhysRevD.92.026009} {\bibfield  {journal} {\bibinfo  {journal} {Phys.
  Rev.}\ }\textbf {\bibinfo {volume} {D92}},\ \bibinfo {pages} {026009}
  (\bibinfo {year} {2015})},\ \Eprint {http://arxiv.org/abs/1505.05012}
  {arXiv:1505.05012 [hep-th]} \BibitemShut {NoStop}%
\bibitem [{\citenamefont {Sybesma}\ and\ \citenamefont
  {Vandoren}(2015)}]{Sybesma:2015oha}%
  \BibitemOpen
  \bibfield  {author} {\bibinfo {author} {\bibfnamefont {Watse}\ \bibnamefont
  {Sybesma}}\ and\ \bibinfo {author} {\bibfnamefont {Stefan}\ \bibnamefont
  {Vandoren}},\ }\bibfield  {title} {\enquote {\bibinfo {title} {{Lifshitz
  quasinormal modes and relaxation from holography}},}\ }\href {\doibase
  10.1007/JHEP05(2015)021} {\bibfield  {journal} {\bibinfo  {journal} {JHEP}\
  }\textbf {\bibinfo {volume} {05}},\ \bibinfo {pages} {021} (\bibinfo {year}
  {2015})},\ \Eprint {http://arxiv.org/abs/1503.07457} {arXiv:1503.07457
  [hep-th]} \BibitemShut {NoStop}%
\bibitem [{\citenamefont {Kovtun}(2012)}]{Kovtun:2012rj}%
  \BibitemOpen
  \bibfield  {author} {\bibinfo {author} {\bibfnamefont {Pavel}\ \bibnamefont
  {Kovtun}},\ }\bibfield  {title} {\enquote {\bibinfo {title} {{Lectures on
  hydrodynamic fluctuations in relativistic theories}},}\ }\bibfield
  {booktitle} {\emph {\bibinfo {booktitle} {{INT Summer School on Applications
  of String Theory Seattle, Washington, USA, July 18-29, 2011}}},\ }\href
  {\doibase 10.1088/1751-8113/45/47/473001} {\bibfield  {journal} {\bibinfo
  {journal} {J. Phys.}\ }\textbf {\bibinfo {volume} {A45}},\ \bibinfo {pages}
  {473001} (\bibinfo {year} {2012})},\ \Eprint {http://arxiv.org/abs/1205.5040}
  {arXiv:1205.5040 [hep-th]} \BibitemShut {NoStop}%
\bibitem [{\citenamefont {Roberts}\ and\ \citenamefont
  {Swingle}(2016)}]{Roberts:2016wdl}%
  \BibitemOpen
  \bibfield  {author} {\bibinfo {author} {\bibfnamefont {Daniel~A.}\
  \bibnamefont {Roberts}}\ and\ \bibinfo {author} {\bibfnamefont {Brian}\
  \bibnamefont {Swingle}},\ }\bibfield  {title} {\enquote {\bibinfo {title}
  {{Lieb-Robinson Bound and the Butterfly Effect in Quantum Field Theories}},}\
  }\href {\doibase 10.1103/PhysRevLett.117.091602} {\bibfield  {journal}
  {\bibinfo  {journal} {Phys. Rev. Lett.}\ }\textbf {\bibinfo {volume} {117}},\
  \bibinfo {pages} {091602} (\bibinfo {year} {2016})},\ \Eprint
  {http://arxiv.org/abs/1603.09298} {arXiv:1603.09298 [hep-th]} \BibitemShut
  {NoStop}%
\bibitem [{\citenamefont {{Karch}}\ \emph {et~al.}(2009)\citenamefont
  {{Karch}}, \citenamefont {{Son}},\ and\ \citenamefont
  {{Starinets}}}]{Karch:2008fa}%
  \BibitemOpen
  \bibfield  {author} {\bibinfo {author} {\bibfnamefont {A.}~\bibnamefont
  {{Karch}}}, \bibinfo {author} {\bibfnamefont {D.~T.}\ \bibnamefont {{Son}}},
  \ and\ \bibinfo {author} {\bibfnamefont {A.~O.}\ \bibnamefont
  {{Starinets}}},\ }\bibfield  {title} {\enquote {\bibinfo {title} {{Zero Sound
  from Holography}},}\ }\href {\doibase 10.1103/PhysRevLett.102.051602}
  {\bibfield  {journal} {\bibinfo  {journal} {Phys. Rev. Lett.}\ }\textbf
  {\bibinfo {volume} {102}},\ \bibinfo {pages} {051602} (\bibinfo {year}
  {2009})},\ \Eprint {http://arxiv.org/abs/0806.3796} {arXiv:0806.3796
  [hep-th]} \BibitemShut {NoStop}%
\bibitem [{\citenamefont {Hartnoll}\ \emph {et~al.}(2010)\citenamefont
  {Hartnoll}, \citenamefont {Polchinski}, \citenamefont {Silverstein},\ and\
  \citenamefont {Tong}}]{Hartnoll:2009ns}%
  \BibitemOpen
  \bibfield  {author} {\bibinfo {author} {\bibfnamefont {Sean~A.}\ \bibnamefont
  {Hartnoll}}, \bibinfo {author} {\bibfnamefont {Joseph}\ \bibnamefont
  {Polchinski}}, \bibinfo {author} {\bibfnamefont {Eva}\ \bibnamefont
  {Silverstein}}, \ and\ \bibinfo {author} {\bibfnamefont {David}\ \bibnamefont
  {Tong}},\ }\bibfield  {title} {\enquote {\bibinfo {title} {{Towards strange
  metallic holography}},}\ }\href {\doibase 10.1007/JHEP04(2010)120} {\bibfield
   {journal} {\bibinfo  {journal} {JHEP}\ }\textbf {\bibinfo {volume} {04}},\
  \bibinfo {pages} {120} (\bibinfo {year} {2010})},\ \Eprint
  {http://arxiv.org/abs/0912.1061} {arXiv:0912.1061 [hep-th]} \BibitemShut
  {NoStop}%
\bibitem [{\citenamefont {Davison}\ and\ \citenamefont
  {Starinets}(2012)}]{Davison:2011ek}%
  \BibitemOpen
  \bibfield  {author} {\bibinfo {author} {\bibfnamefont {Richard~A.}\
  \bibnamefont {Davison}}\ and\ \bibinfo {author} {\bibfnamefont {Andrei~O.}\
  \bibnamefont {Starinets}},\ }\bibfield  {title} {\enquote {\bibinfo {title}
  {{Holographic zero sound at finite temperature}},}\ }\href {\doibase
  10.1103/PhysRevD.85.026004} {\bibfield  {journal} {\bibinfo  {journal} {Phys.
  Rev.}\ }\textbf {\bibinfo {volume} {D85}},\ \bibinfo {pages} {026004}
  (\bibinfo {year} {2012})},\ \Eprint {http://arxiv.org/abs/1109.6343}
  {arXiv:1109.6343 [hep-th]} \BibitemShut {NoStop}%
\bibitem [{\citenamefont {Witczak-Krempa}(2014)}]{Witczak-Krempa:2013aea}%
  \BibitemOpen
  \bibfield  {author} {\bibinfo {author} {\bibfnamefont {William}\ \bibnamefont
  {Witczak-Krempa}},\ }\bibfield  {title} {\enquote {\bibinfo {title} {{Quantum
  critical charge response from higher derivatives in holography}},}\ }\href
  {\doibase 10.1103/PhysRevB.89.161114} {\bibfield  {journal} {\bibinfo
  {journal} {Phys. Rev.}\ }\textbf {\bibinfo {volume} {B89}},\ \bibinfo {pages}
  {161114} (\bibinfo {year} {2014})},\ \Eprint {http://arxiv.org/abs/1312.3334}
  {arXiv:1312.3334 [cond-mat.str-el]} \BibitemShut {NoStop}%
\bibitem [{\citenamefont {Davison}\ and\ \citenamefont
  {Gout\'eraux}(2015{\natexlab{b}})}]{Davison:2014lua}%
  \BibitemOpen
  \bibfield  {author} {\bibinfo {author} {\bibfnamefont {Richard~A.}\
  \bibnamefont {Davison}}\ and\ \bibinfo {author} {\bibfnamefont {Blaise}\
  \bibnamefont {Gout\'eraux}},\ }\bibfield  {title} {\enquote {\bibinfo {title}
  {{Momentum dissipation and effective theories of coherent and incoherent
  transport}},}\ }\href {\doibase 10.1007/JHEP01(2015)039} {\bibfield
  {journal} {\bibinfo  {journal} {JHEP}\ }\textbf {\bibinfo {volume} {1501}},\
  \bibinfo {pages} {039} (\bibinfo {year} {2015}{\natexlab{b}})},\ \Eprint
  {http://arxiv.org/abs/1411.1062} {arXiv:1411.1062 [hep-th]} \BibitemShut
  {NoStop}%
\bibitem [{\citenamefont {Grozdanov}\ \emph {et~al.}(2016)\citenamefont
  {Grozdanov}, \citenamefont {Kaplis},\ and\ \citenamefont
  {Starinets}}]{Grozdanov:2016vgg}%
  \BibitemOpen
  \bibfield  {author} {\bibinfo {author} {\bibfnamefont {Sa\v{s}o}\
  \bibnamefont {Grozdanov}}, \bibinfo {author} {\bibfnamefont {Nikolaos}\
  \bibnamefont {Kaplis}}, \ and\ \bibinfo {author} {\bibfnamefont {Andrei~O.}\
  \bibnamefont {Starinets}},\ }\bibfield  {title} {\enquote {\bibinfo {title}
  {{From strong to weak coupling in holographic models of thermalization}},}\
  }\href {\doibase 10.1007/JHEP07(2016)151} {\bibfield  {journal} {\bibinfo
  {journal} {JHEP}\ }\textbf {\bibinfo {volume} {07}},\ \bibinfo {pages} {151}
  (\bibinfo {year} {2016})},\ \Eprint {http://arxiv.org/abs/1605.02173}
  {arXiv:1605.02173 [hep-th]} \BibitemShut {NoStop}%
\bibitem [{\citenamefont {Grozdanov}\ and\ \citenamefont
  {Starinets}(2017)}]{Grozdanov:2016fkt}%
  \BibitemOpen
  \bibfield  {author} {\bibinfo {author} {\bibfnamefont {Sa\v{s}o}\
  \bibnamefont {Grozdanov}}\ and\ \bibinfo {author} {\bibfnamefont {Andrei~O.}\
  \bibnamefont {Starinets}},\ }\bibfield  {title} {\enquote {\bibinfo {title}
  {{Second-order transport, quasinormal modes and zero-viscosity limit in the
  Gauss-Bonnet holographic fluid}},}\ }\href {\doibase 10.1007/JHEP03(2017)166}
  {\bibfield  {journal} {\bibinfo  {journal} {JHEP}\ }\textbf {\bibinfo
  {volume} {03}},\ \bibinfo {pages} {166} (\bibinfo {year} {2017})},\ \Eprint
  {http://arxiv.org/abs/1611.07053} {arXiv:1611.07053 [hep-th]} \BibitemShut
  {NoStop}%
\bibitem [{\citenamefont {Chen}\ and\ \citenamefont
  {Lucas}(2017)}]{Chen:2017dsy}%
  \BibitemOpen
  \bibfield  {author} {\bibinfo {author} {\bibfnamefont {Chi-Fang}\
  \bibnamefont {Chen}}\ and\ \bibinfo {author} {\bibfnamefont {Andrew}\
  \bibnamefont {Lucas}},\ }\bibfield  {title} {\enquote {\bibinfo {title}
  {{Origin of the Drude peak and of zero sound in probe brane holography}},}\
  }\href {\doibase 10.1016/j.physletb.2017.10.023} {\bibfield  {journal}
  {\bibinfo  {journal} {Phys. Lett.}\ }\textbf {\bibinfo {volume} {B774}},\
  \bibinfo {pages} {569--574} (\bibinfo {year} {2017})},\ \Eprint
  {http://arxiv.org/abs/1709.01520} {arXiv:1709.01520 [hep-th]} \BibitemShut
  {NoStop}%
\bibitem [{\citenamefont {Grozdanov}\ \emph {et~al.}(2019)\citenamefont
  {Grozdanov}, \citenamefont {Lucas},\ and\ \citenamefont
  {Poovuttikul}}]{Grozdanov:2018fic}%
  \BibitemOpen
  \bibfield  {author} {\bibinfo {author} {\bibfnamefont {Sašo}\ \bibnamefont
  {Grozdanov}}, \bibinfo {author} {\bibfnamefont {Andrew}\ \bibnamefont
  {Lucas}}, \ and\ \bibinfo {author} {\bibfnamefont {Napat}\ \bibnamefont
  {Poovuttikul}},\ }\bibfield  {title} {\enquote {\bibinfo {title} {{Holography
  and hydrodynamics with weakly broken symmetries}},}\ }\href {\doibase
  10.1103/PhysRevD.99.086012} {\bibfield  {journal} {\bibinfo  {journal} {Phys.
  Rev.}\ }\textbf {\bibinfo {volume} {D99}},\ \bibinfo {pages} {086012}
  (\bibinfo {year} {2019})},\ \Eprint {http://arxiv.org/abs/1810.10016}
  {arXiv:1810.10016 [hep-th]} \BibitemShut {NoStop}%
\bibitem [{\citenamefont {{Kadanoff}}\ and\ \citenamefont
  {{Martin}}(1963)}]{1963AnPhy..24..419K}%
  \BibitemOpen
  \bibfield  {author} {\bibinfo {author} {\bibfnamefont {L.~P.}\ \bibnamefont
  {{Kadanoff}}}\ and\ \bibinfo {author} {\bibfnamefont {P.~C.}\ \bibnamefont
  {{Martin}}},\ }\bibfield  {title} {\enquote {\bibinfo {title} {{Hydrodynamic
  equations and correlation functions}},}\ }\href {\doibase
  10.1016/0003-4916(63)90078-2} {\bibfield  {journal} {\bibinfo  {journal}
  {Annals of Physics}\ }\textbf {\bibinfo {volume} {24}},\ \bibinfo {pages}
  {419--469} (\bibinfo {year} {1963})}\BibitemShut {NoStop}%
\bibitem [{\citenamefont {Hartnoll}\ \emph {et~al.}(2007)\citenamefont
  {Hartnoll}, \citenamefont {Kovtun}, \citenamefont {Muller},\ and\
  \citenamefont {Sachdev}}]{Hartnoll:2007ih}%
  \BibitemOpen
  \bibfield  {author} {\bibinfo {author} {\bibfnamefont {Sean~A.}\ \bibnamefont
  {Hartnoll}}, \bibinfo {author} {\bibfnamefont {Pavel~K.}\ \bibnamefont
  {Kovtun}}, \bibinfo {author} {\bibfnamefont {Markus}\ \bibnamefont {Muller}},
  \ and\ \bibinfo {author} {\bibfnamefont {Subir}\ \bibnamefont {Sachdev}},\
  }\bibfield  {title} {\enquote {\bibinfo {title} {{Theory of the Nernst effect
  near quantum phase transitions in condensed matter, and in dyonic black
  holes}},}\ }\href {\doibase 10.1103/PhysRevB.76.144502} {\bibfield  {journal}
  {\bibinfo  {journal} {Phys.Rev.}\ }\textbf {\bibinfo {volume} {B76}},\
  \bibinfo {pages} {144502} (\bibinfo {year} {2007})},\ \Eprint
  {http://arxiv.org/abs/0706.3215} {arXiv:0706.3215 [cond-mat.str-el]}
  \BibitemShut {NoStop}%
\bibitem [{\citenamefont {Davison}\ \emph {et~al.}()\citenamefont {Davison},
  \citenamefont {Gout\'eraux},\ and\ \citenamefont
  {Poovuttikul}}]{Davison2019}%
  \BibitemOpen
  \bibfield  {author} {\bibinfo {author} {\bibfnamefont {Richard~A.}\
  \bibnamefont {Davison}}, \bibinfo {author} {\bibfnamefont {Blaise}\
  \bibnamefont {Gout\'eraux}}, \ and\ \bibinfo {author} {\bibfnamefont {Napat}\
  \bibnamefont {Poovuttikul}},\ }\href@noop {} {\ }\bibinfo {note} {In
  progress}\BibitemShut {NoStop}%
\bibitem [{\citenamefont {Phillips}\ and\ \citenamefont
  {Chamon}(2005)}]{PhysRevLett.95.107002}%
  \BibitemOpen
  \bibfield  {author} {\bibinfo {author} {\bibfnamefont {Philip}\ \bibnamefont
  {Phillips}}\ and\ \bibinfo {author} {\bibfnamefont {Claudio}\ \bibnamefont
  {Chamon}},\ }\bibfield  {title} {\enquote {\bibinfo {title} {Breakdown of
  one-parameter scaling in quantum critical scenarios for high-temperature
  copper-oxide superconductors},}\ }\href {\doibase
  10.1103/PhysRevLett.95.107002} {\bibfield  {journal} {\bibinfo  {journal}
  {Phys. Rev. Lett.}\ }\textbf {\bibinfo {volume} {95}},\ \bibinfo {pages}
  {107002} (\bibinfo {year} {2005})}\BibitemShut {NoStop}%
\end{thebibliography}%

\end{document}